\documentclass[preprint, 12pt]{elsarticle}
\usepackage[margin = 0.8in]{geometry}
\usepackage{amsmath, amssymb}
\usepackage{array}
\usepackage{graphicx}
\usepackage{epstopdf}
\ifpdf
\DeclareGraphicsExtensions{.eps,.pdf,.png,.jpg}
\else
  \DeclareGraphicsExtensions{.eps}
\fi
\usepackage{enumitem}
\usepackage{booktabs}

\usepackage{hyperref} 
\usepackage[capitalise]{cleveref}

\allowdisplaybreaks

\newcommand{\W}{\textit{Wolbachia}\,\,}
\newcommand{\Wns}{\textit{Wolbachia}}
\usepackage{tikz}
\DeclareRobustCommand*\circled[1]{\tikz[baseline=(char.base)]{\node[shape=circle,draw,inner sep=1pt] (char) {#1};}}

\begin{document}

\begin{frontmatter}
\title{Structures of the Basic Reproduction Number $\mathcal{R}_0$ Across Compartmental Disease Models}
\author[1]{Zhuolin Qu\corref{cor1}}
\ead{zhuolin.qu@utsa.edu}
\author[2,3]{Abhi Ashwath}
\ead{abhi.ashwath@berkeley.edu}

\cortext[cor1]{Corresponding author}

\affiliation[1]{organization={Department of Mathematics, University of Texas at San Antonio},
city={San Antonio},
state={TX}}

\affiliation[2]{organization={Department of Mathematics, University of California, Berkeley, Berkeley, CA}}
\affiliation[3]{organization={Saint Mary's Hall},
city={San Antonio},
state= {TX}}

\begin{abstract}
The basic reproduction number $\mathcal{R}_0$ is the central dimensionless quantity in mathematical epidemiology, characterizing the threshold for disease outbreak and the early growth rate of an epidemic. The algebraic form of $\mathcal{R}_0$ varies widely across models of distinct transmission mechanisms, and its interpretation can yield further biological insight into transmission dynamics and disease intervention. We present a structural taxonomy of $\mathcal{R}_0$ for compartmental ordinary differential equation models spanning a range of disease transmission features, including staged progression, differential infectivity, competing strains, vector-borne transmission, sexually transmitted infection, recurrent infection, maternal transmission, and gene drive inheritance. For each structural class, we provide a worked example, deriving $\mathcal{R}_0$ via the next-generation matrix approach and interpreting the resulting algebraic expression in terms of the underlying transmission pathway. By unifying disparate derivations under a single structural framework, we clarify when and why particular algebraic forms of $\mathcal{R}_0$ arise. This review serves both as a research synthesis and as a self-contained pedagogical reference for students and researchers entering the field.
\end{abstract}


\end{frontmatter}

\section{Introduction}\label{sec:intro}
The basic reproduction number, $\mathcal{R}_0$, is the central dimensionless quantity in mathematical epidemiology, with origins tracing back to the foundational work of Kermack and McKendrick \cite{kermack1927contribution}. Formally, $\mathcal{R}_0$ is defined as the expected number of secondary infections produced by one infected individual introduced into an entirely susceptible population \cite{diekmann1990definition}. This corresponds to a threshold condition: an epidemic can occur if and only if $\mathcal{R}_0>1$, and its value governs the initial rate of spread once the invasion occurs \cite{heffernan2005perspectives}. The most widely used method for computing $\mathcal{R}_0$ in compartmental models is the next-generation matrix (NGM) approach, first introduced by Diekmann et al.\ \cite{diekmann1990definition} and later formalized for ordinary differential equation compartmental models by van den Driessche and Watmough \cite{van2002reproduction}; this approach decomposes the infected subsystem into new-infection and transition components and defines $\mathcal{R}_0$ as the spectral radius of the NGM. 

Although $\mathcal{R}_0$ is most often treated purely as a threshold value, its algebraic structure offers rich insight into the underlying transmission mechanism. Whether $\mathcal{R}_0$ takes the form of a weighted sum, a maximum, a geometric series, a geometric mean, or a ratio is not incidental: each form reflects a distinct qualitative feature of how infection moves through the population, and recognizing this structure is itself part of interpreting $\mathcal{R}_0$ correctly.

In this review, we survey a range of $\mathcal{R}_0$ structures across a variety of compartmental models, organized into two groups: horizontal transmission, in which infection passes between infected and susceptible individuals within a generation, and vertical transmission, in which infection passes from parent to offspring through reproduction \cite{busenberg1993vertically}. The resulting taxonomy, summarized in \cref{tab:taxonomy}, is intended to serve two purposes: as a research synthesis, organizing existing $\mathcal{R}_0$ results under a structural taxonomy, and as a pedagogical reference, in the spirit of \cite{blackwood2018compartmental}, offering readers new to the field a structured entry point into $\mathcal{R}_0$ derivations and interpretations across a range of transmission mechanisms.

The paper is organized as follows. In \cref{sec:NGM}, we provide a brief walkthrough of the NGM process for computing $\mathcal{R}_0$ using a simple SEIR model. In \cref{sec:horizontal} and \cref{sec:vertical}, we survey models from horizontal and vertical transmission mechanisms, respectively. In \cref{sec:conclusion}, we discuss further biological insights offered by the $\mathcal{R}_0$ taxonomy.

\begin{table}[htbp]
\centering
\caption{Structural forms of $\mathcal{R}_0$ surveyed.}
\label{tab:taxonomy}
\setlength{\tabcolsep}{3pt}
\begin{tabular}{p{3.2cm}p{4cm}>{\raggedright\arraybackslash}p{6cm}p{1.5cm}}
\toprule
\textbf{Structure} & \textbf{Algebraic form} & \textbf{Biological mechanism} & \textbf{Section} \\
\midrule
Weighted sum & $\displaystyle\sum_i p_i\,\mathcal{R}_0^{(i)}$
& Heterogeneous infectivity across substages, weighted by entry probability
& Sec.\ \ref{sec:DISP} \\ \midrule
Maximum & $\displaystyle\max_i\{\mathcal{R}_0^{(i)}\}$
& Multiple strains/patches, non-interacting but sharing the same host pool, dominated by the fastest-growing one
& Sec.\ \ref{sec:multimax} \\\midrule
Geometric series (cyclic) & $\displaystyle\mathcal{R}_0^{\mathrm{cycle}}/(1-P_\gamma) \newline = \mathcal{R}_0^{\mathrm{cycle}} \sum_{i=0}^\infty {(P_\gamma)^i}$
& Repeated relapse cycles, each rescaled by the relapse probability $P_\gamma$
& Sec.\ \ref{sec:cyclic} \\\midrule
Geometric mean & $\sqrt{\mathcal{R}_{HV}\,\mathcal{R}_{VH}}$
& Bipartite transmission cycle (vector-borne or heterosexual STD)
& Sec.\ \ref{sec:bipartite} \\\midrule
Fitness ratio & $\mathcal{G}_{0w}/\mathcal{G}_{0u}$
& Vertical/genetic transmission as competition between resident and introduced cohorts
& Sec.\ \ref{sec:wolbachia} \& \ref{sec:medea}\\
\bottomrule
\end{tabular}
\end{table}

\section{A Brief Introduction to the Next-Generation Matrix Method for Computing $\mathcal{R}_0$}\label{sec:NGM}

The basic reproduction number, $\mathcal{R}_0$, serves as the threshold governing local stability of the disease-free equilibrium (DFE).
For simple models such as SIR model, this threshold can be read off directly from the sign of $dI/dt$ at the DFE; for models with multiple infected stages, the same threshold is more systematically obtained via linearization of the infected subsystem, a classical argument that we do not repeat here; we instead refer the reader to standard texts such as \cite{hethcote2000mathematics, martcheva2015introduction,brauer2019mathematical}. Throughout this paper, we derive $\mathcal{R}_0$ using the NGM method \cite{diekmann1990definition,van2002reproduction}, and we demonstrate the procedure with the classical SEIR model.

\begin{figure}[htbp]
\centering
\includegraphics[width=0.6\linewidth]{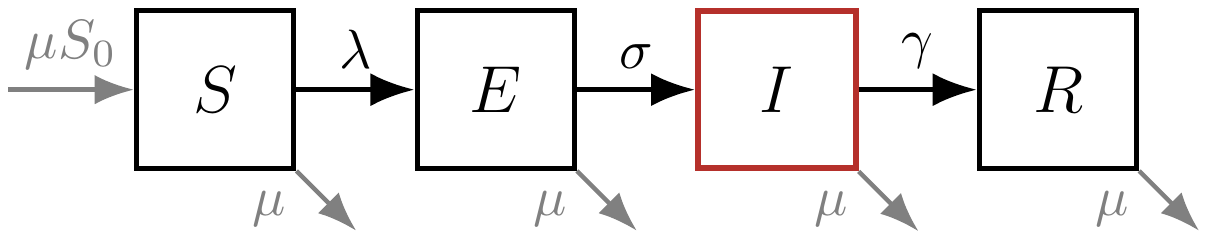}
\caption{Schematic representation of the SEIR model with demography.}
\label{fig:SEIR}
\end{figure}

We consider the SEIR model (\cref{fig:SEIR}), in which individuals progress through susceptible ($S$), exposed ($E$), infectious ($I$), and recovered ($R$) stages; exposed individuals have been infected but are not yet infectious, and progress to the infectious stage at per-capita rate $\sigma$, while infectious individuals recover at per-capita rate $\gamma$. With births and deaths occurring at the constant per-capita rate $\mu$, the total population $N=S+E+I+R$ remains constant at its disease-free value $S_0$. The governing system is
\begin{align}
\frac{dS}{dt} &= \mu(S_0-S) - \lambda S , \nonumber\\
\frac{dE}{dt} &= \lambda S - (\sigma+\mu)E, \label{eq:seir}\\
\frac{dI}{dt} &= \sigma E - (\gamma+\mu)I, \nonumber\\
\frac{dR}{dt} &= \gamma I - \mu R, \nonumber
\end{align}
with force of infection $\lambda(I) = c\beta I/N$, where $c$ is the average contact rate per person per unit time and $\beta$ is the per-contact transmission probability. The unique DFE is $(S,E,I,R)^\ast = (S_0,0,0,0)$.

\paragraph{$\mathcal{F}$--$\mathcal{V}$ decomposition} Following the NGM approach, we take the infected state vector to be $\mathbf{x}=(E,I)^\top$ and separate new infections, $\mathcal{F}(\mathbf{x})$, from transitions, $\mathcal{V}(\mathbf{x})$, for the corresponding infected subsystem:
\begin{equation*}
\frac{d\mathbf{x}}{dt} = \mathcal{F}(\mathbf{x})-\mathcal{V}(\mathbf{x}),\quad \text{where}\quad 
\mathcal{F} = \begin{pmatrix} c\beta S \dfrac{I}{N} \\[6pt] 0 \end{pmatrix}, \quad \mathcal{V} = \begin{pmatrix} (\sigma+\mu)E \\[4pt] -\sigma E + (\gamma+\mu)I \end{pmatrix}.
\end{equation*}
Evaluating the Jacobians of $\mathcal{F}(\mathbf{x})$ and $\mathcal{V}(\mathbf{x})$ at the DFE, where $S=N=S_0$,
\begin{equation*}
F = \frac{\partial \mathcal{F}}{\partial \mathbf{x}} \bigg| _{\text{DFE}}=\begin{pmatrix} 0 & c\beta \\ 0 & 0 \end{pmatrix}, \qquad V = \frac{\partial \mathcal{V}}{\partial \mathbf{x}} \bigg| _{\text{DFE}} = \begin{pmatrix} \sigma+\mu & 0 \\ -\sigma & \gamma+\mu \end{pmatrix},
\end{equation*}
and the resulting NGM is given by
\begin{equation*}
FV^{-1} = \begin{pmatrix} \dfrac{c\beta\sigma}{(\sigma+\mu)(\gamma+\mu)} & \dfrac{c\beta}{\gamma+\mu} \\[8pt] 0 & 0 \end{pmatrix}.
\end{equation*}
\paragraph{Basic reproduction number} Since $FV^{-1}$ is a triangular matrix, the spectral radius (the maximum of the absolute values of its eigenvalues) is simply the nonzero $(1,1)$ entry, giving
\begin{equation}\label{eq:R0_SEIR}
\mathcal{R}_0 = \rho(FV^{-1}) = \frac{c\beta\sigma}{(\sigma+\mu)(\gamma+\mu)}.
\end{equation}
This construction process can be generalized to arbitrary-dimensional infected subsystems.

\paragraph{$\mathcal{R}_0$ structure and disease transmission mechanism} 
The resulting expression \cref{eq:R0_SEIR} can be interpreted as a product of three factors: $\sigma/(\sigma+\mu)$ is the probability that an individual in $E$ stage survives the incubation period to reach the  $I$ stage (not lost to natural mortality $\mu$), $c\beta$ is the transmission rate per unit time in the $I$ stage (contacts $c$ per unit time, each transmitting with probability $\beta$), and $1/(\gamma+\mu)$ is the average time spent in the $I$ stage. The product of these three factors recovers exactly the definition of $\mathcal{R}_0$: the expected number of secondary infections that an infected individual can generate over its infectious period, in an otherwise fully susceptible population.

One caveat applies to the interpretation of $\mathcal{R}_0$ as a disease threshold. As established rigorously in \cite{van2002reproduction}, the correspondence between $\mathcal{R}_0$ and DFE stability is strictly local. It guarantees only that a sufficiently small introduction of infected individuals near the DFE will either take off or not, with no claim about global dynamics far from the DFE. In fact, as detailed in \cite{van2002reproduction}, some models may admit an unstable sub-threshold endemic equilibrium when $\mathcal{R}_0<1$, leading to a subcritical bifurcation (or backward bifurcation), where initial conditions above the unstable endemic equilibrium lead to sustained disease prevalence even though $\mathcal{R}_0<1$. 

% -----------------------------------------------------------------------
\section{Horizontal Transmission}
\label{sec:horizontal}

Horizontal transmission refers to the spread of infection between individuals (or between an individual and a vector) within the same generation, in contrast to vertical transmission, which occurs from parents to offspring. The subsections below include a range of horizontal transmission with different features, progressing from simple multi-stage infections through increasingly complex population and transmission structures.

\subsection{Staged Progression and Differential Infectivity: $\mathcal{R}_0$ as Weighted Sum} \label{sec:DISP}
Many infectious diseases exhibit variation in infectiousness that either develops progressively over time within an individual (such as acute, chronic, and late phases) or is a fixed characteristic across individuals in the population (such as children and adults). We consider two such models with multiple infection groups, the staged progression (SP) model and the differential infectivity (DI) model. Both frameworks are developed in \cite{hyman1999differential}, where HIV serves as the motivating application.

\begin{figure}[htbp]
\centering
\includegraphics[width=0.7\linewidth]{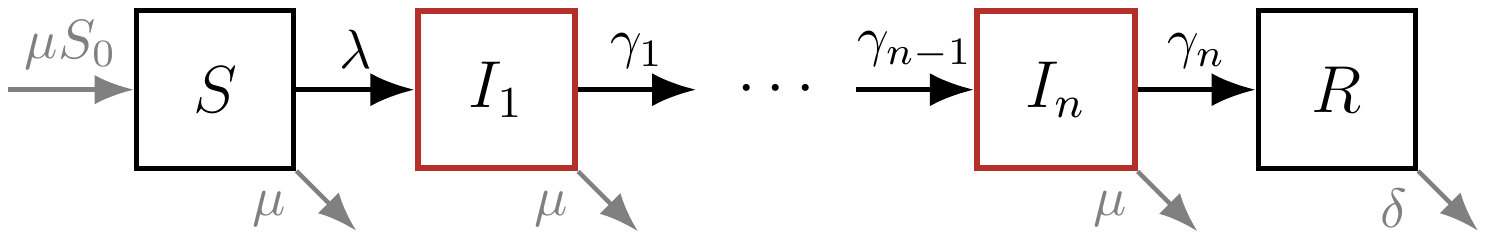}
\includegraphics[width=0.5\linewidth]{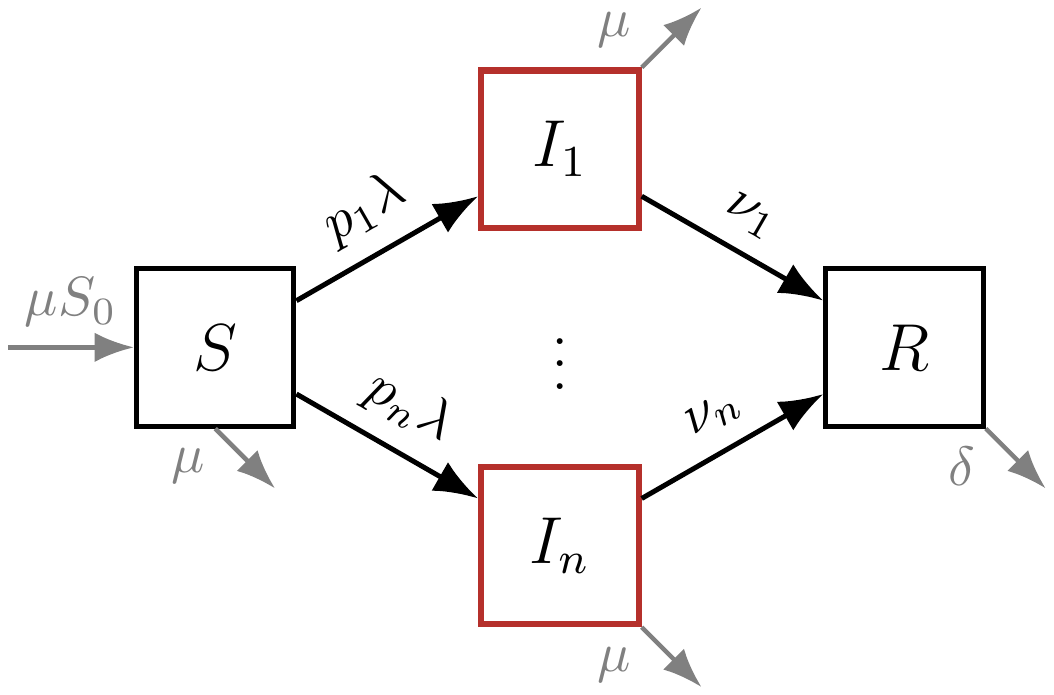}
\caption{Schematic representation for the staged-progression model (top), in which individuals advance sequentially through infection stages $I_1, I_2, \cdots, I_n$, and the differential-infectivity model (bottom), in which individuals are assigned at infection to one of n parallel groups.}
\label{fig:SP_DI}
\end{figure}

In the SP model, all newly infected individuals enter stage $I_1$ and advance sequentially through $n$ infection stages $I_1\to I_2\to\cdots\to I_n$, capturing the temporal progression of viral load or clinical severity (\cref{fig:SP_DI}, Top). In the DI model, the infected population is partitioned at the moment of infection into $n$ mutually exclusive subgroups $I_1,\dots,I_n$ with distinct transmission rates, reflecting stable biological heterogeneity rather than temporal progression (\cref{fig:SP_DI}, Bottom). 

\textbf{SP model.} All infected individuals enter stage $I_1$ and progress sequentially through $n$ stages at rates $\gamma_1,\dots,\gamma_n$:
\begin{align*}
\frac{dS}{dt}   &= \mu(S_0-S) - \lambda S, \\
\frac{dI_1}{dt} &= \lambda S - (\gamma_1+\mu)I_1, \\
\frac{dI_i}{dt} &= \gamma_{i-1}I_{i-1} - (\gamma_i+\mu)I_i, \quad 2\le i\le n,\\
\frac{dR}{dt} &= \gamma_n I_n -\delta R,
\end{align*}
where the force of infection is given by 
\begin{equation}\label{eq:SPDI_FOI}
\lambda = c\sum_{i=1}^n\beta_i \frac{I_i}{N},
\end{equation}
$c$ is the average contact rate per person per unit time, $\beta_i$ is the transmission probability for the $I_i$ stage, and $N = S+\sum_{i=1}^n I_i$, assuming removed individuals in the $R$ stage no longer interact with the rest of the population. $S_0$ is the stable susceptible population size in the absence of infection, $\mu$ is the baseline birth and mortality rate, and $\delta\ge \mu$ accounts for additional disease-induced mortality.

\textbf{DI model.} The infected population is divided into $n$ subgroups $I_1,\dots,I_n$. Upon infection, an individual enters group $I_i$ with probability $p_i$ $(\sum_{i=1}^n p_i=1)$, where each group has its own transmission probability $\beta_i$ and removal rate $\nu_i$:
\begin{align*}
\frac{dS}{dt}   &= \mu(S_0-S) - \lambda S, \\
\frac{dI_i}{dt} &= p_i\lambda S - (\nu_i+\mu)I_i, \quad i=1,\dots,n,\\
\frac{dR}{dt}   &= \sum_{i=1}^n \nu_i I_i - \delta R,
\end{align*}
where the force of infection, $\lambda$, is the same as in the SP model, given in \cref{eq:SPDI_FOI}, and the other parameters are the same as in the SP model.

\paragraph{$\mathcal{F}$--$\mathcal{V}$ decomposition} For both models, the infected state vector is $\mathbf{x}=(I_1,\dots,I_n)^\top$, and the DFE is $(S^*,I_1^*,\dots,I_n^*,R^*) = (S_0,0,\dots,0,0)$.

\underline{SP model.}
\begin{equation*}
\mathcal{F}^{SP}
= \begin{pmatrix} 
\lambda S \\ 
0 \\ 
\vdots \\ 
0 
\end{pmatrix}, \qquad 
\mathcal{V}^{SP} = 
\begin{pmatrix} 
(\gamma_1+\mu)I_1 \\ 
- \gamma_1 I_1+ (\gamma_2+\mu)I_2  \\ 
\vdots \\ 
- \gamma_{n-1}I_{n-1}+ (\gamma_n+\mu)I_n  
\end{pmatrix}.
\end{equation*}
Evaluating the Jacobians at the DFE,
\begin{equation*}
F^{SP} = \begin{pmatrix} c\beta_1 & c\beta_2 & \cdots & c\beta_n \\ 0 & 0 & \cdots & 0 \\ \vdots & & \ddots & \vdots \\ 0 & 0 & \cdots & 0 \end{pmatrix},\qquad
V^{SP} = \begin{pmatrix}
\gamma_1+\mu &         &        &         \\
-\gamma_1    & \gamma_2+\mu &        &         \\
& \ddots  & \ddots &         \\
&         & -\gamma_{n-1} & \gamma_n+\mu
\end{pmatrix},
\end{equation*}
and the NGM is
\begin{equation}\label{eq:FV_SP}
F^{SP}(V^{SP})^{-1} = 
\begin{pmatrix} 
r_1 & r_2 & \cdots & r_n \\ 
0 & 0 & \cdots & 0 \\ 
\vdots & & \ddots & \vdots \\ 
0 & 0 & \cdots & 0 
\end{pmatrix}, \qquad 
r_j = c\sum_{k=j}^{n}\frac{\beta_k\displaystyle\prod_{l=j}^{k-1}\gamma_l}{\displaystyle\prod_{l=j}^{k}(\gamma_l+\mu)}.
\end{equation}

\underline{DI model.}
\begin{equation*}
\mathcal{F}^{DI} = 
\begin{pmatrix} 
p_1\lambda S \\ 
p_2\lambda S \\ 
\vdots \\ 
p_n\lambda S 
\end{pmatrix}, \qquad 
\mathcal{V}^{DI} = 
\begin{pmatrix} 
(\nu_1+\mu)I_1 \\ 
(\nu_2+\mu)I_2 \\ 
\vdots \\ 
(\nu_n+\mu)I_n
\end{pmatrix}.
\end{equation*}
Evaluating the Jacobians at the DFE,
\begin{equation*}
  F^{DI} = \begin{pmatrix}
    p_1 c\beta_1 & p_1 c\beta_2 & \cdots & p_1 c\beta_n \\
    p_2 c\beta_1 & p_2 c\beta_2 & \cdots & p_2 c\beta_n \\
    \vdots       &              & \ddots & \vdots       \\
    p_n c\beta_1 & \cdots       &        & p_n c\beta_n
  \end{pmatrix},
  \qquad
  V^{DI} = \begin{pmatrix}
    \nu_1+\mu & & \\
    & \ddots & \\
    & & \nu_n+\mu
  \end{pmatrix},
\end{equation*}
and the NGM is
\begin{equation}\label{eq:FV_DI}
F^{DI}(V^{DI})^{-1} = c
\begin{pmatrix} 
\dfrac{p_1\beta_1}{\nu_1+\mu} & \dfrac{p_1\beta_2}{\nu_2+\mu} & \cdots & \dfrac{p_1\beta_n}{\nu_n+\mu} \\[6pt] 
\dfrac{p_2\beta_1}{\nu_1+\mu} & \dfrac{p_2\beta_2}{\nu_2+\mu} & \cdots & \dfrac{p_2\beta_n}{\nu_n+\mu} \\ 
\vdots & & \ddots & \vdots \\ 
\dfrac{p_n\beta_1}{\nu_1+\mu} & \dfrac{p_n\beta_2}{\nu_2+\mu} & \cdots & \dfrac{p_n\beta_n}{\nu_n+\mu} 
\end{pmatrix}.
\end{equation}

\paragraph{Basic reproduction number} Taking the spectral radius of the next-generation matrices for both models yields the basic reproduction numbers.

\underline{SP model.} From \cref{eq:FV_SP}, $F^{SP}(V^{SP})^{-1}$ is upper triangular, so its eigenvalues are its diagonal entries $(r_1, 0, \dots, 0)$, giving
\begin{equation}\label{eq:R0_SP}
\mathcal{R}_0^{SP} = \rho(F^{SP}(V^{SP})^{-1}) = r_1 = \sum_{k=1}^{n}\underbrace{\left(\prod_{l=1}^{k-1}\frac{\gamma_l}{\gamma_l+\mu}\right)}_{\text{prob. of reaching stage $k$}}
\overbrace{\frac{c\beta_k}{\gamma_k+\mu}}^{\hspace{-2cm}\text{contact $\times$ trans. $\times$ duration}\hspace{-2cm}}.
\end{equation}

\underline{DI model.} From \cref{eq:FV_DI}, $F^{DI}(V^{DI})^{-1}$ has rank one; its unique nonzero eigenvalue equals its trace:
\begin{equation}\label{eq:R0_DI}
\mathcal{R}_0^{DI} = \rho(F^{DI}(V^{DI})^{-1}) = \sum_{k=1}^n \underbrace{p_k}_{\text{prob. of entering stage $k$}}\hspace{-1cm}  \overbrace{\frac{c\beta_k}{\nu_k+\mu}}^{\hspace{-2cm}\text{contact $\times$ trans. $\times$ duration}\hspace{-2cm}}.
\end{equation}

\paragraph{$\mathcal{R}_0$ structure and disease transmission mechanism} The basic reproduction numbers for both models admit a unified representation as a weighted sum of group-level reproduction numbers,
\begin{equation*}
\mathcal{R}_0 = \sum_{k=1}^n \mathcal{P}_k\,\mathcal{R}_0^{(k)},
\end{equation*}
where $\mathcal{P}_k$ is the probability that a newly infected individual reaches stage $k$ and $\mathcal{R}_0^{(k)}$ is the expected number of secondary infections produced by a single individual in stage $k$ in an otherwise fully susceptible population. The weighted-sum form is not specific to either SP or DI model; it is the signature of any model in which new infections enter through a single shared force-of-infection channel and are subsequently routed into non-interacting stages, so that each stage's contribution, $\mathcal{R}_0^{(k)}$, to the global transmission is generated independently, weighted by the probability of reaching that stage, $\mathcal{P}_k$. 

In the DI model, $\mathcal{P}_k = p_k$ directly, since group membership is assigned at the moment of infection. In the SP model, all individuals begin in $I_1$ (so $\mathcal{P}_1=1$), and the probability of surviving through all preceding stages to reach $I_k$ is $\mathcal{P}_k = \prod_{l=1}^{k-1}\gamma_l/(\gamma_l+\mu)$. The group-level reproduction number $\mathcal{R}_0^{(k)} = c\beta_k/(\text{exit rate from } I_k)$ shares a common form for both models: the product of the contact rate $c$, the per-contact transmission probability $\beta_k$, and the mean exit time for group $k$. The SP and DI models thus represent two complementary forms of infectivity heterogeneity, sequential and parallel, respectively, that yield structurally identical expressions for $\mathcal{R}_0$, differing only in the entry probabilities $\mathcal{P}_k$.

\subsection{Competing Strain: $\mathcal{R}_0$ as a Maximum}
\label{sec:multimax}
When multiple strains of a pathogen circulate simultaneously in a population, transmission preserves strain identity: an infected individual's secondary cases inherit the same strain, so the strains compete only indirectly, through their shared draw on the pool of susceptible hosts needed to persist in the population. This structure is relevant whenever the distinguishing feature between groups is a heritable property carried forward by the pathogen through transmission. Typical applications include competing variants of the same pathogen, as in strain-replacement dynamics, and drug-sensitive versus drug-resistant strains. The same structure may arise, in a degenerate form, in multi-patch models with no movement between patches: new infections occur strictly locally within each patch, and the heritable property is location. 

This competing-strain structure differs fundamentally from that of the DI model introduced in \cref{sec:DISP}. Although both models distribute new infections among multiple infectivity groups, the multi-strain model carries a heritable feature that the DI model lacks: the origin of a transmission event determines its destination, so that infections generated by group $i$ can only produce new cases in group $i$. In the DI model, by contrast, new infections from all groups are pooled into a single force of infection and then redistributed across groups according to fixed proportions that depend on characteristics of the recipient, such as age, immune status, and prior exposure, rather than on which group transmitted the infection.

\begin{figure}[htbp]
\centering
\includegraphics[width=0.5\linewidth]{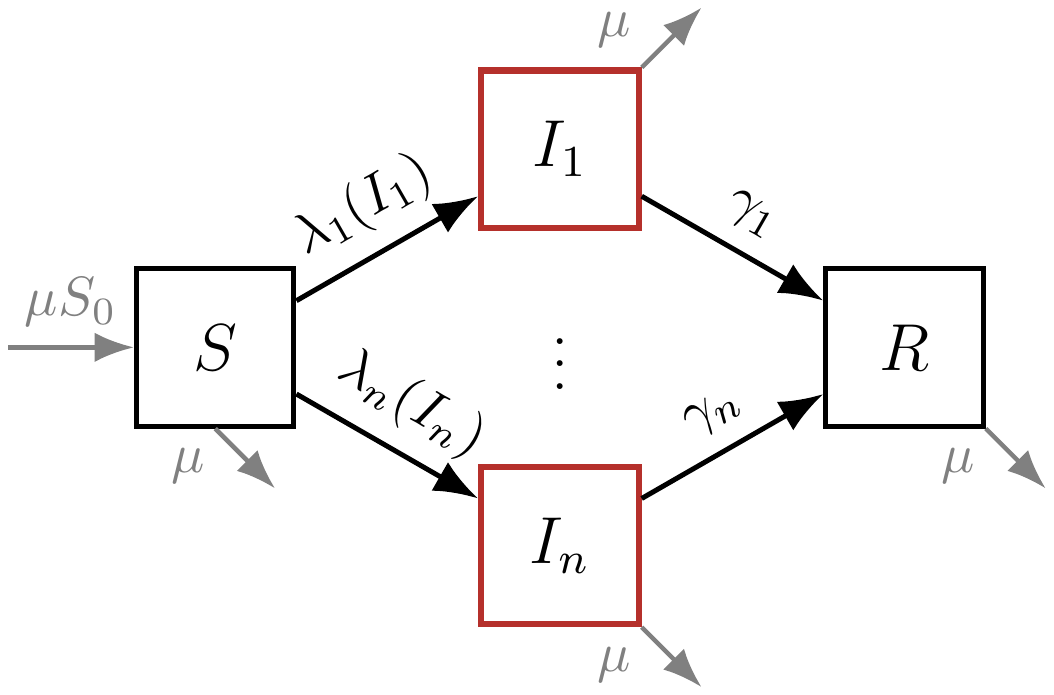}
\caption{Schematic representation of the multi-strain model: each infected group $I_i$ transmits only to its own infected group via strain-specific force of infection $\lambda_i(I_i)$.}
\label{fig:ms}
\end{figure}

We consider a similar SIR-type model with multiple infection groups as in the DI model. Each infection group $I_i$ has its own transmission rate $\beta_i$ and recovery rate $\gamma_i$, and we assume that transmission preserves strain identity, i.e., a susceptible individual infected by strain $i$ always enters stage $I_i$ and no other. In a closed population of size $N = S + \sum_i I_i + R$, the dynamics are given by 
\begin{align*} 
\frac{dS}{dt} &= \mu(S_0-S) - \lambda S, \\ 
\frac{dI_i}{dt} &= \lambda_i S - (\gamma_i+\mu) I_i, \qquad i = 1,\dots,n, \\ 
\frac{dR}{dt} &= \sum_{i=1}^n \gamma_i I_i- \mu R, 
\end{align*}
where the strain-specific force of infection, $\lambda_i $, and total force of infection, $\lambda$, are given by
\begin{equation*}
\lambda_i = c \beta_i \frac{I_i}{N},\quad \text{and}\quad \lambda = \sum_{i=1}^n \lambda_i.
\end{equation*}
The unique DFE is $(S,I_1,\dots,I_n,R)^\ast = (S_0,0,\dots,0,0)$. 

\paragraph{$\mathcal{F}$--$\mathcal{V}$ decomposition} With infected state vector $\mathbf{x} = (I_1,\dots,I_n)^\top$, the new-infection and transition vectors are
\begin{equation*}
\mathcal{F} = \begin{pmatrix} c\beta_1 \dfrac{I_1}{N}S \\[6pt] \vdots \\[4pt] c\beta_n \dfrac{I_n}{N}S \end{pmatrix}, \qquad \mathcal{V} = \begin{pmatrix} (\gamma_1+\mu)I_1 \\ \vdots \\ (\gamma_n+\mu)I_n \end{pmatrix}.
\end{equation*}
Because transmission preserves strain identity, $\mathcal{F}_i$ depends only on $I_i$ and not on any $I_j$, $j \neq i$; consequently, the Jacobians evaluated at the DFE are diagonal, $F = \mathrm{diag}(c\beta_1,\dots,c\beta_n)$ and $V = \mathrm{diag}(\gamma_1+\mu,\dots,\gamma_n+\mu)$, and the NGM is
\begin{equation*}
FV^{-1} = \mathrm{diag}\!\left(\frac{c\beta_1}{\gamma_1+\mu},\,\dots,\,\frac{c\beta_n}{\gamma_n+\mu}\right).
\end{equation*}

\paragraph{Basic reproduction number} Since $FV^{-1}$ is diagonal, the spectral radius follows immediately
\begin{equation}\label{eq:R0_max}
\mathcal{R}_0 = \rho(FV^{-1}) = \max_{1\le i\le n}\mathcal{R}_0^{(i)}, \qquad \mathcal{R}_0^{(i)} = \frac{c\beta_i}{\gamma_i+\mu}.
\end{equation}

\paragraph{$\mathcal{R}_0$ structure and disease transmission mechanism} The diagonal NGM reflects a structural fact about phenotype-preserving transmission: because strain identity is a heritable property of the pathogen, an infected individual's secondary cases always belong to the same strain, so that no strain's early growth depends on the abundance of any other. In contrast, in the SP or DI model, different infection stages contribute to and are affected by the global force of infection. Here, the linearized system decouples into $n$ separate one-dimensional growth problems, each governed by its own strain-specific $\mathcal{R}_0^{(i)}$. The global $\mathcal{R}_0$ is then the maximum of the strain-level $\mathcal{R}_0^{(i)}$, that is, the global dynamics are dominated by the single most transmissible strain, which can deplete the shared susceptible pool the fastest \cite{bremermann1989competitivea}.

\subsection{Cyclic Reproduction Number for Disease Relapse} \label{sec:cyclic}
Many infectious diseases permit repeated symptomatic episodes within a single infection course. Rather than clearing an infection following primary illness, individuals may enter a persistent asymptomatic carrier state from which they subsequently relapse to active disease. This cyclical pattern characterizes \emph{Plasmodium vivax} malaria, in which hepatic hypnozoites reactivate to produce recurrent blood-stage infection \cite{white2014modelling}; invasive non-typhoidal \emph{Salmonella} (iNTS), in which gastrointestinal carriage precedes episodes of recurring invasive disease \cite{qu2021staged}; and tuberculosis, in which latent bacillary persistence drives reactivation disease \cite{blower1995intrinsic}. In each case, a single host may cycle through symptomatic and asymptomatic phases multiple times before recovering fully or dying, and each symptomatic episode contributes to onward transmission. $\mathcal{R}_0$ must therefore account for the full sequence of potential relapse bouts, leading to the form of a geometric series.

\begin{figure}[htbp]
\centering
\includegraphics[width=0.45\linewidth]{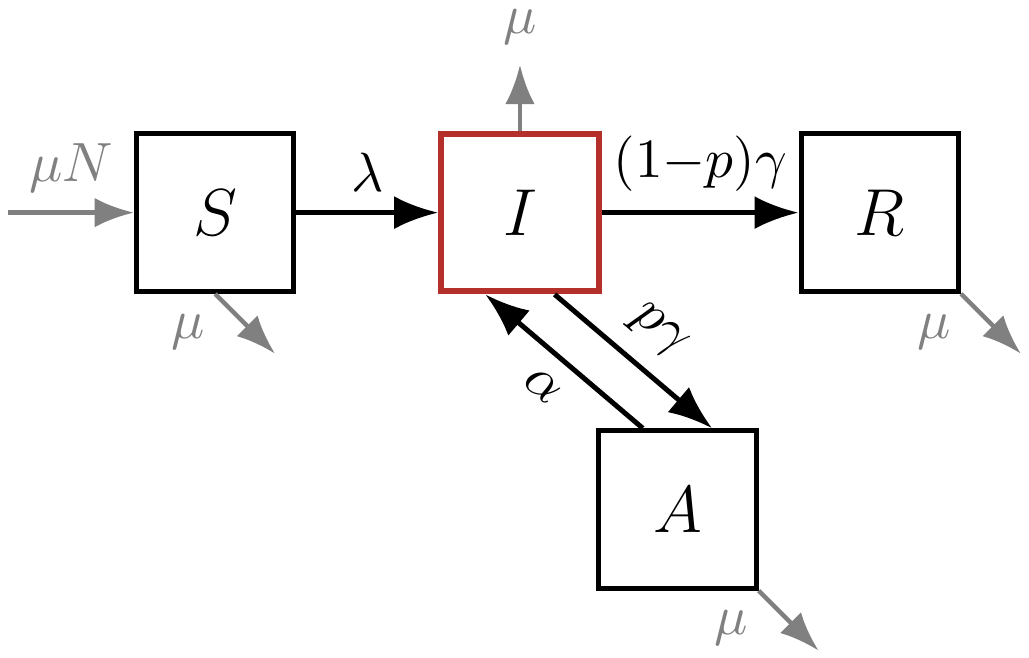}
\caption{Schematic representation of the SIAR model with disease relapse between asymptomatic infected stage $A$ and symptomatic infected stage $I$.}
\end{figure}

We consider a minimal SIAR model comprising susceptible ($S$), symptomatic infected ($I$), asymptomatic infected ($A$), and fully recovered ($R$) stages, with constant total population $N$ and uniform per-capita birth and natural death rate $\mu$. Symptomatic individuals exit $I$ and progress to the next disease stage at total per-capita rate $\gamma$: a fraction $p$ progresses to the asymptomatic stage at rate $p\gamma$, and a fraction $1-p$ recovers permanently to $R$ at rate $(1-p)\gamma$. Asymptomatic individuals may relapse to $I$ at rate $\alpha$. Assuming only symptomatic individuals contribute to transmission, the force of infection is $\lambda = c\beta I/N$, and the governing system is
\begin{align}
\frac{dS}{dt} &= \mu N - \lambda S - \mu S, \nonumber\\
\frac{dI}{dt} &= \lambda S - (\gamma + \mu)I + \alpha A, \label{eq:I_siar}\\
\frac{dA}{dt} &= p\gamma\, I - (\alpha + \mu)A, \nonumber\\
\frac{dR}{dt} &= (1-p)\gamma\, I - \mu R. \nonumber
\end{align}
The relapse term $\alpha A$ in \cref{eq:I_siar} represents reactivation of a pre-existing infection rather than a new transmission event; it is therefore classified as a transition rather than a new infection in the $\mathcal{F}$--$\mathcal{V}$ decomposition below. The unique DFE is $(S,I,A,R)^\ast = (N,0,0,0)$.

\paragraph{$\mathcal{F}$--$\mathcal{V}$ decomposition} 
With infected state vector $\mathbf{x} = (I,A)^\top$, the new-infection and transition vectors are
\begin{equation*}
\mathcal{F} = \begin{pmatrix} c\beta\dfrac{I}{N}S \\[8pt] 0 \end{pmatrix}, \qquad \mathcal{V} = \begin{pmatrix} (\gamma+\mu)I - \alpha A \\[6pt] -p\gamma\,I + (\alpha+\mu)A \end{pmatrix}.
\end{equation*}
The Jacobians at the DFE are
\begin{equation*}
F = \begin{pmatrix} c\beta & 0 \\ 0 & 0 \end{pmatrix}, \qquad V = \begin{pmatrix} \gamma+\mu & -\alpha \\ -p\gamma & \alpha+\mu \end{pmatrix},
\end{equation*}
and the NGM is
\begin{equation}\label{eq:NGM_siar}
FV^{-1} = \frac{1}{\det V}\begin{pmatrix} c\beta(\alpha+\mu) & c\beta\alpha \\ 0 & 0 \end{pmatrix}, \qquad \det V = (\gamma+\mu)(\alpha+\mu) - p\gamma\alpha.
\end{equation}

\paragraph{Basic reproduction number} 
Since $FV^{-1}$ is upper triangular with a zero second row, its spectral radius equals the $(1,1)$ entry, giving
\begin{equation}\label{eq:R0_siar_raw}
\mathcal{R}_0 = \rho(FV^{-1}) = \frac{c\beta(\alpha+\mu)}{(\gamma+\mu)(\alpha+\mu) - p\gamma\alpha}.
\end{equation}

\paragraph{$\mathcal{R}_0$ interpretation without demography}  To interpret \cref{eq:R0_siar_raw}, we first set $\mu = 0$ to isolate the transmission structure from demographic turnover, which reduces the expression to
\begin{equation}\label{eq:R0_siar_mu0}
\mathcal{R}_0\big|_{\mu=0} = \frac{c\beta}{\gamma(1-p)}.
\end{equation}
This result admits a direct probabilistic derivation. During a single symptomatic episode, one infectious individual generates new infections at rate $c\beta$ for an expected duration $1/\gamma$, producing $c\beta/\gamma$ secondary cases. At the conclusion of that episode, the individual enters the asymptomatic stage with probability $p$ or recovers permanently with probability $1-p$. In the absence of demography, any individual reaching $A$ relapses to $I$, so the probability of reaching the $(k+1)$-th symptomatic episode is exactly $p^k$. The expected secondary infections generated during that episode are therefore $(c\beta/\gamma)\,p^k$. Summing over all possible bouts yields the \emph{cyclic reproduction number}
\begin{equation}\label{eq:R0_siar}
\mathcal{R}_0 = \sum_{k=0}^{\infty} \frac{c\beta}{\gamma}\,p^k = \frac{c\beta}{\gamma} \cdot \frac{1}{1-p},
\end{equation}
in agreement with \cref{eq:R0_siar_mu0}.

The geometric series structure in \cref{eq:R0_siar} generalizes to a broader class of relapse models. Denoting by $\mathcal{R}_0^{\mathrm{cycle}}$ the expected new infections produced during any single symptomatic episode and by $P_\gamma$ the probability of surviving one complete cycle to initiate the next, the cyclic reproduction number takes the general form
\begin{equation}\label{eq:cycle}
\mathcal{R}_0 = \sum_{k=0}^{\infty} \mathcal{R}_0^{\mathrm{cycle}}\,(P_\gamma)^k = \frac{\mathcal{R}_0^{\mathrm{cycle}}}{1 - P_\gamma},
\end{equation}
assuming $P_\gamma < 1$. The factor $1/(1-P_\gamma)$ equals the expected total number of symptomatic episodes experienced by a single infected individual, so $\mathcal{R}_0$ can be read as the per-episode reproduction number amplified by that expected episode count.

\paragraph{$\mathcal{R}_0$ interpretation with demography}  Returning to general $\mu > 0$, expression \cref{eq:R0_siar_raw} can be rewritten to match the cyclic form \cref{eq:cycle}:
\begin{equation*}
\mathcal{R}_0 = \frac{c\beta(\alpha+\mu)}{(\gamma+\mu)(\alpha+\mu) - p\gamma\alpha} = \frac{c\beta}{\gamma+\mu} \cdot \frac{1}{1 - \dfrac{p\gamma}{\gamma+\mu}\cdot\dfrac{\alpha}{\alpha+\mu}},
\end{equation*}
identifying
\begin{equation*}
\mathcal{R}_0^{\mathrm{cycle}} = \frac{c\beta}{\gamma+\mu}, \qquad P_\gamma = \frac{p\gamma}{\gamma+\mu} \cdot \frac{\alpha}{\alpha+\mu}.
\end{equation*}
The per-cycle reproduction number $\mathcal{R}_0^{\mathrm{cycle}} = c\beta/(\gamma+\mu)$ reflects the expected infection period in stage $I$ per cycle, $1/(\gamma+\mu)$. The cycle probability $P_\gamma$ decomposes as the product of two conditional survival factors: $p\gamma/(\gamma+\mu)$, the probability of exiting the infectious stage $I$ by progressing to $A$ rather than by dying or recovering permanently, and $\alpha/(\alpha+\mu)$, the probability of surviving the asymptomatic period to relapse back to $I$.

\paragraph{$\mathcal{R}_0$ structure and disease transmission mechanism} 
The geometric series in \cref{eq:cycle} partitions $\mathcal{R}_0$ into contributions from successive symptomatic episodes: the $k$-th term, $\mathcal{R}_0^{\mathrm{cycle}}\,P_\gamma^k$, is the expected secondary infections generated during the $(k+1)$-th episode, weighted by $P_\gamma^k$, the probability of surviving to that episode. This infinite sum arises from the fact that reactivation from the asymptomatic stage is a transition rather than a new infection, and the cyclic form is a general signature of any transmission mechanism permitting infection relapses.

Additionally, since $\mathcal{R}_0>\mathcal{R}_0^{\mathrm{cycle}}$ for $P_\gamma>0$, diseases with high relapse probability can sustain transmission even when the per-episode reproduction number, $\mathcal{R}_0^{\mathrm{cycle}}$, lies substantially below unity. 

\subsection{Example of Combined Structures} \label{sec:salmonella}
In practice, disease models may be a combination of multiple structures. Here we present one example that shows how multiple structural features can be reflected in a single $\mathcal{R}_0$.

Invasive nontyphoidal \textit{Salmonella} (iNTS) causes severe bloodstream infection in sub-Saharan Africa; the circulating ST-313 lineage is human-adapted, with sustained person-to-person transmission. The burden is concentrated in children under five and adults aged 25--40. In both cohorts, there are asymptomatic carriers, who may later relapse to active disease. The model proposed by Qu et al.\ \cite{qu2021staged} combines multiple structures introduced previously: (1) within each risk group, infection advances through a sequence of disease stages (\cref{sec:DISP}, SP model); (2) across episodes, asymptomatic carriage and recurrent infection lead to infection loops (\cref{sec:cyclic}); and (3) across risk groups, new infections are generated based on the susceptibility of the recipient group (\cref{sec:DISP}, DI model). The resulting $\mathcal{R}_0$ reflects all three features.

\begin{figure}[htbp]
\centering
\includegraphics[width=0.9\linewidth]{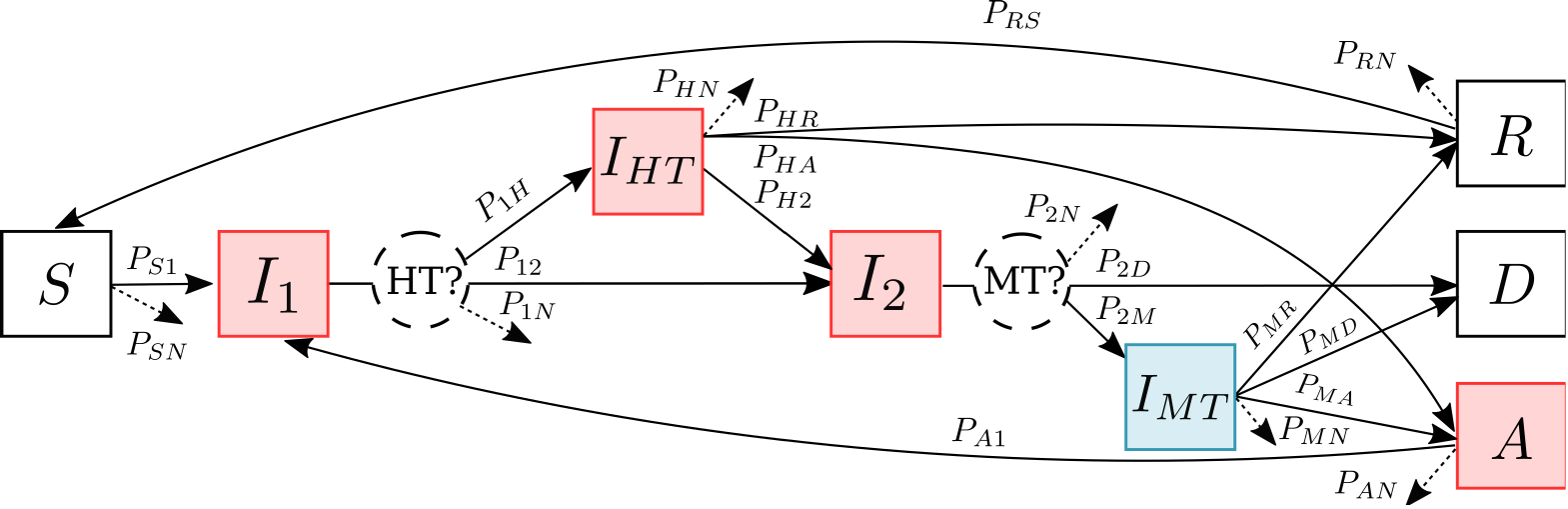}
\caption{Schematic representation of disease progression, represented as a branching process, for iNTS. Branching probability $P_{jk}$ is the fraction of people who progress from stage $j$ to stage $k$, and the progression rates in the model are defined using the branching fractions $P_{jk}$ and the time spent in the source stage $\tau_j$, as defined in \cref{eq:branching}. Note that the risk group index $\kappa$ is suppressed for clarity.}
\label{fig:NTS}
\end{figure}

Four risk groups are tracked, indexed by $\kappa = 1,\dots, 4$. Within each group, people move through infection stages depending on their treatment outcomes (\cref{fig:NTS}): susceptible ($S^\kappa$), mild early infection ($I_1^\kappa$), mild infection under home treatment ($I_{\scriptscriptstyle HT}^\kappa$), severe acute infection ($I_2^\kappa$), severe infection under medical treatment ($I_{\scriptscriptstyle MT}^\kappa$, isolated and non-infectious), asymptomatic carriage ($A^\kappa$, infectious and capable of relapsing to $I_1^\kappa$), temporary immunity ($R^\kappa$), and disease-induced death ($D^\kappa$). For ease of parameterization and interpretation, the per-capita rate of movement from stage $j$ to stage $k$ within group $\kappa$, $\gamma_{jk}^\kappa$, is defined as
\begin{equation}\label{eq:branching}
\gamma_{jk}^\kappa = P_{jk}^\kappa/\tau_j^\kappa,   
\end{equation}
where $P_{jk}^\kappa$ is the branching probability, the fraction of people who progress from stage $j$ to stage $k$, and $\tau_j^\kappa$ is the mean stage duration for stage $j$. The governing system is
\begin{align*}
\frac{dS^\kappa}{dt} &= \mu^\kappa\bigl(S_0^\kappa - S^\kappa\bigr) - \lambda^\kappa S^\kappa + \gamma_{RS}^\kappa R^\kappa,
& \frac{dI_{\scriptscriptstyle MT}^\kappa}{dt} &= \gamma_{\scriptscriptstyle 2M}^\kappa I_2^\kappa - \bigl(\gamma_{\scriptscriptstyle MR}^\kappa + \gamma_{\scriptscriptstyle MD}^\kappa + \gamma_{\scriptscriptstyle MA}^\kappa + \mu^\kappa\bigr) I_{\scriptscriptstyle MT}^\kappa, \\
\frac{dI_1^\kappa}{dt} &= \lambda^\kappa S^\kappa + \gamma_{\scriptscriptstyle A1}^\kappa A^\kappa - \bigl(\gamma_{\scriptscriptstyle 1H}^\kappa + \gamma_{12}^\kappa + \mu^\kappa\bigr) I_1^\kappa,
& \frac{dA^\kappa}{dt} &= \gamma_{\scriptscriptstyle HA}^\kappa I_{\scriptscriptstyle HT}^\kappa + \gamma_{\scriptscriptstyle MA}^\kappa I_{\scriptscriptstyle MT}^\kappa - \bigl(\gamma_{\scriptscriptstyle A1}^\kappa + \mu^\kappa\bigr) A^\kappa, \\
\frac{dI_{\scriptscriptstyle HT}^\kappa}{dt} &= \gamma_{\scriptscriptstyle 1H}^\kappa I_1^\kappa - \bigl(\gamma_{\scriptscriptstyle HR}^\kappa + \gamma_{\scriptscriptstyle HA}^\kappa + \gamma_{\scriptscriptstyle H2}^\kappa + \mu^\kappa\bigr) I_{\scriptscriptstyle HT}^\kappa,
& \frac{dR^\kappa}{dt} &= \gamma_{\scriptscriptstyle HR}^\kappa I_{\scriptscriptstyle HT}^\kappa + \gamma_{\scriptscriptstyle MR}^\kappa I_{\scriptscriptstyle MT}^\kappa - (\gamma_{RS}^\kappa+\mu^\kappa)R^\kappa, \\
\frac{dI_2^\kappa}{dt} &= \gamma_{12}^\kappa I_1^\kappa + \gamma_{\scriptscriptstyle H2}^\kappa I_{\scriptscriptstyle HT}^\kappa - \bigl(\gamma_{\scriptscriptstyle 2D}^\kappa + \gamma_{\scriptscriptstyle 2M}^\kappa + \mu^\kappa\bigr) I_2^\kappa,
& \frac{dD^\kappa}{dt} &= \gamma_{\scriptscriptstyle 2D}^\kappa I_2^\kappa + \gamma_{\scriptscriptstyle MD}^\kappa I_{\scriptscriptstyle MT}^\kappa,
\end{align*}
for $\kappa=1,\dots,4$. The groups are coupled only through the force of infection,
\begin{equation}\label{eq:iNTS_FOI}
\lambda^\kappa = c\beta^\kappa\,
\frac{\sum_{\ell=1}^{4}\bigl(I_1^\ell + I_{\scriptscriptstyle HT}^\ell + I_2^\ell + A^\ell\bigr)}{\sum_{\ell=1}^{4}\bigl(S^\ell + I_1^\ell + I_{\scriptscriptstyle HT}^\ell + I_2^\ell + A^\ell + R^\ell\bigr)},
\end{equation}
where $c$ is the contact rate, assumed equal across groups, and $\beta^\kappa$ is the group-specific transmission probability. Medically treated individuals are assumed isolated, so $I_{\scriptscriptstyle MT}$ is absent from both the denominator and numerator of \cref{eq:iNTS_FOI}.

\paragraph{$\mathcal{F}$--$\mathcal{V}$ decomposition} Ordering the infected states by group as $\mathbf{x} = (\mathbf{x}^1,\dots,\mathbf{x}^4)^\top$ with $\mathbf{x}^\kappa = (I_1^\kappa, I_{\scriptscriptstyle HT}^\kappa, I_2^\kappa, I_{\scriptscriptstyle MT}^\kappa, A^\kappa)^\top$, and the new infection and transition vectors are $\mathcal{F}:=(\mathcal{F}^1;\mathcal{F}^2;\mathcal{F}^3;\mathcal{F}^4)$ and $\mathcal{V}:=(\mathcal{V}^1;\mathcal{V}^2;\mathcal{V}^3;\mathcal{V}^4)$, 
where
\begin{equation*}
\mathcal{F}^\kappa = \bigl(\lambda^\kappa S^\kappa,\;0,\;0,\;0,\;0\bigr)^\top,
\qquad
\mathcal{V}^\kappa =
\begin{pmatrix}
\bigl(\gamma_{\scriptscriptstyle 1H}^\kappa + \gamma_{12}^\kappa + \mu^\kappa\bigr) I_1^\kappa - \gamma_{\scriptscriptstyle A1}^\kappa A^\kappa \\[4pt]
\bigl(\gamma_{\scriptscriptstyle HR}^\kappa + \gamma_{\scriptscriptstyle HA}^\kappa + \gamma_{\scriptscriptstyle H2}^\kappa + \mu^\kappa\bigr) I_{\scriptscriptstyle HT}^\kappa - \gamma_{\scriptscriptstyle 1H}^\kappa I_1^\kappa \\[4pt]
\bigl(\gamma_{\scriptscriptstyle 2D}^\kappa + \gamma_{\scriptscriptstyle 2M}^\kappa + \mu^\kappa\bigr) I_2^\kappa - \gamma_{12}^\kappa I_1^\kappa - \gamma_{\scriptscriptstyle H2}^\kappa I_{\scriptscriptstyle HT}^\kappa \\[4pt]
\bigl(\gamma_{\scriptscriptstyle MR}^\kappa + \gamma_{\scriptscriptstyle MD}^\kappa + \gamma_{\scriptscriptstyle MA}^\kappa + \mu^\kappa\bigr) I_{\scriptscriptstyle MT}^\kappa - \gamma_{\scriptscriptstyle 2M}^\kappa I_2^\kappa \\[4pt]
\bigl(\gamma_{\scriptscriptstyle A1}^\kappa + \mu^\kappa\bigr) A^\kappa - \gamma_{\scriptscriptstyle HA}^\kappa I_{\scriptscriptstyle HT}^\kappa - \gamma_{\scriptscriptstyle MA}^\kappa I_{\scriptscriptstyle MT}^\kappa
\end{pmatrix}.
\end{equation*}
The DFE is $(S^\kappa,I_1^\kappa,I_{\scriptscriptstyle HT}^\kappa,I_2^\kappa,I_{\scriptscriptstyle MT}^\kappa,A^\kappa,R^\kappa)^\ast = (S_0^\kappa,0,0,0,0,0,0)$ for each $\kappa$, with $N_0 = \sum_{\ell=1}^4 S_0^\ell$. The explicit Jacobian matrices are given in Appendix \ref{sec:app_nts}. 

\paragraph{Basic reproduction number}
After algebraic simplifications, the  basic reproduction number is given by
\begin{equation}\label{eq:R0_NTS} 
\mathcal{R}_0 = \sum_{\kappa=1}^4 \frac{S_0^\kappa}{N_0}\,\mathcal{R}_0^\kappa,
\end{equation}
where
\begin{equation*}
\mathcal{R}_0^\kappa=c\beta^{\kappa}\frac{\tau_1^{\kappa}+P_{\scriptscriptstyle 1H}^{\kappa}\tau_H^{\kappa}+\big(P_{12}^{\kappa}+P_{\scriptscriptstyle 1H}^{\kappa}P_{\scriptscriptstyle H2}^{\kappa}\big)\tau_2^{\kappa}+\big(P_{\scriptscriptstyle 1H}^{\kappa}P_{\scriptscriptstyle HA}^{\kappa}+P_{\scriptscriptstyle 1H}^{\kappa}P_{\scriptscriptstyle H2}^{\kappa}P_{\scriptscriptstyle 2M}^{\kappa}P_{\scriptscriptstyle MA}^{\kappa}+P_{12}^{\kappa}P_{\scriptscriptstyle 2M}^{\kappa}P_{\scriptscriptstyle MA}^{\kappa}\big)\tau_A^{\kappa}}{1-P_{\scriptscriptstyle A1}^{\kappa}\big(P_{\scriptscriptstyle 1H}^{\kappa}P_{\scriptscriptstyle HA}^{\kappa}+P_{\scriptscriptstyle 1H}^{\kappa}P_{\scriptscriptstyle H2}^{\kappa}P_{\scriptscriptstyle 2M}^{\kappa}P_{\scriptscriptstyle MA}^{\kappa}+P_{12}^{\kappa}P_{\scriptscriptstyle 2M}^{\kappa}P_{\scriptscriptstyle MA}^{\kappa}\big)}.
\end{equation*}
The basic reproduction number expression in \cref{eq:R0_NTS} reflects the three structural features of the model:
\begin{enumerate}[leftmargin=1.5em]
\item The staged progression within each risk group enters the numerator of the $\mathcal{R}_0^\kappa$:
\begin{align*}
(\mathcal{R}_0^\kappa)^{\text{cycle}} &:= c \beta^{\kappa}\Bigl(\tau_1^{\kappa}+P_{\scriptscriptstyle 1H}^{\kappa}\tau_H^{\kappa}+\big(P_{12}^{\kappa}+P_{\scriptscriptstyle 1H}^{\kappa}P_{\scriptscriptstyle H2}^{\kappa}\big)\tau_2^{\kappa} \\
&\qquad{\hspace{1cm}}+\big(P_{\scriptscriptstyle 1H}^{\kappa}P_{\scriptscriptstyle HA}^{\kappa}+P_{\scriptscriptstyle 1H}^{\kappa}P_{\scriptscriptstyle H2}^{\kappa}P_{\scriptscriptstyle 2M}^{\kappa}P_{\scriptscriptstyle MA}^{\kappa}+P_{12}^{\kappa}P_{\scriptscriptstyle 2M}^{\kappa}P_{\scriptscriptstyle MA}^{\kappa}\big)\tau_A^{\kappa}\Bigr),
\end{align*}
representing the total contribution of new secondary cases from each of the four infectious stages, $I_1^\kappa$, $I_{\scriptscriptstyle HT}^\kappa$, $I_2^\kappa$ and $A^\kappa$, where people on average spent $\tau_1^\kappa$, $\tau_H^\kappa$, $\tau_2^\kappa$, $\tau_A^\kappa$ amount of time, respectively. The coefficients for $\tau_{\star}^\kappa$ represent the probabilities of transitions from the initial infected stage $I_1^\kappa$ to the corresponding infectious stages. This mirrors the structure of SP model in \cref{eq:R0_SP}.
\item The infection loops due to recurrent infection of the asymptomatic carriage lead to the structure of a cyclic reproduction number as in \cref{eq:cycle}. Rewriting $\mathcal{R}_0^\kappa$ as
\begin{equation*}
\mathcal{R}_0^\kappa = \frac{(\mathcal{R}_0^\kappa)^{\text{cycle}}}{1-P_\gamma^{\kappa}}, \quad 
P_\gamma^{\kappa} = P_{\scriptscriptstyle A1}^{\kappa}\big(P_{\scriptscriptstyle 1H}^{\kappa}P_{\scriptscriptstyle HA}^{\kappa}+P_{\scriptscriptstyle 1H}^{\kappa}P_{\scriptscriptstyle H2}^{\kappa}P_{\scriptscriptstyle 2M}^{\kappa}P_{\scriptscriptstyle MA}^{\kappa}+P_{12}^{\kappa}P_{\scriptscriptstyle 2M}^{\kappa}P_{\scriptscriptstyle MA}^{\kappa}\big),
\end{equation*}
where $P_\gamma^{\kappa}$ corresponds to the probability of relapse, that is, the probability of progressing from the initial infection stage $I_1^{\kappa}$ to the $A^{\kappa}$ stage (the terms in the parentheses), times the probability of returning to $I_1^{\kappa}$, which is $P_{\scriptscriptstyle A1}^\kappa$.
\item The total contribution across the four risk groups takes the weighted-sum form over the group-specific $\mathcal{R}_0^\kappa$, where the probability of reaching $I_1^\kappa$ is approximated by the fraction of the population, $S_0^\kappa/N_0$, at the DFE. This reflects the DI model structure, similar to \cref{eq:R0_DI}.
\end{enumerate}
This composite example illustrates that the structures described in this paper serve as reusable building blocks rather than mutually exclusive model classes. 

\subsection{Bipartite Transmission: $\mathcal{R}_0$ as a Geometric Mean}\label{sec:bipartite}
Some disease transmissions require completing a cycle between two distinct host classes. In vector-borne disease, transmission is mediated by an arthropod vector, typically a biting insect such as a mosquito, that acquires a pathogen from one host and delivers it to another; a complete infection cycle requires passage through both the host and the vector populations, either as host$\to$vector$\to$host or equivalently vector$\to$host$\to$vector. Heterosexual transmission of a sexually transmitted disease (STD) imposes the same bipartite transmission structure: infection passes from males to females and from females to males, but not within either sex. In both settings, neither population can sustain a chain of transmission in isolation, and this shared bipartite cycle structure yields a geometric-mean form for $\mathcal{R}_0$ expression. 

\subsubsection{Vector-borne Disease}
\begin{figure}[htbp]
\centering
\includegraphics[width=0.3\linewidth]{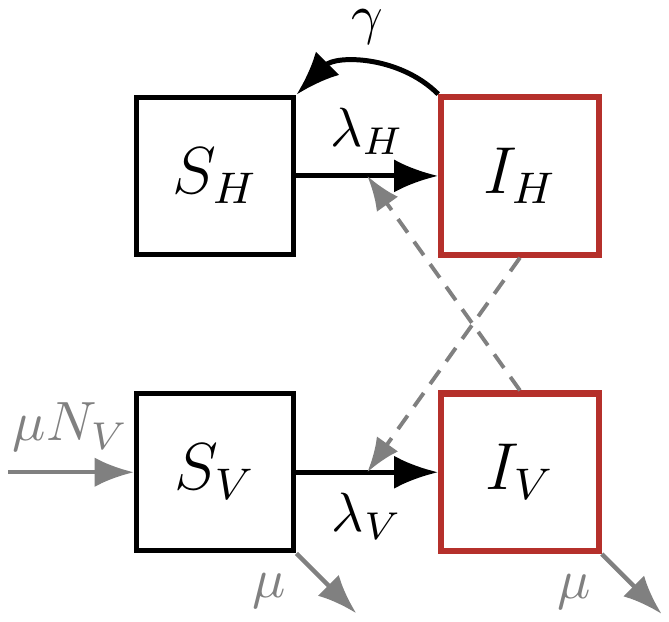}
\caption{Schematic representation of the coupled SIS human and SI vector model for vector-borne disease transmission.}
\label{fig:vector-borne}
\end{figure}

We consider the canonical vector-borne disease model coupling a susceptible-infected-susceptible (SIS) human population with a susceptible-infected (SI) vector population \cite{ross1911prevention, macdonald1957epidemiology}. Let $N_H$ and $N_V$ denote the constant populations sizes of the human and vector, respectively, with $S_H + I_H = N_H$ and $S_V + I_V = N_V$. Humans recover at per-capita rate $\gamma$ and re-enter the susceptible stage; vectors are born susceptible and die at per-capita rate $\mu$, keeping $N_V$ constant; see \cref{fig:vector-borne}.
The governing system is
\begin{align*}
\frac{dS_H}{dt} &= -\lambda_H S_H + \gamma I_H, \qquad \frac{dS_V}{dt} = \mu N_V - \lambda_V S_V - \mu S_V,\\
\frac{dI_H}{dt} &= \lambda_H S_H - \gamma I_H, ~~~\qquad \frac{dI_V}{dt} = \lambda_V S_V - \mu I_V,
\end{align*}
with forces of infection
\begin{equation}\label{eq:foi_vb}
\lambda_H = b_H \beta_{VH} \frac{I_V}{N_V}, \qquad \lambda_V = b_V \beta_{HV} \frac{I_H}{N_H}.
\end{equation}
Here $b_H$ is the per-human biting rate (bites received per human per unit time), $b_V$ is the per-vector biting rate (bites delivered per vector per unit time), $\beta_{VH}$ is the per-bite probability of transmission from an infectious vector to a susceptible human ($V\to H$), and $\beta_{HV}$ is the per-bite probability of transmission from an infectious human to a susceptible vector ($H\to V$). The unique DFE is $(S_H, I_H, S_V, I_V)^\ast = (N_H, 0, N_V, 0)$. 

\paragraph{$\mathcal{F}$--$\mathcal{V}$ decomposition} Taking the infected state vector $\mathbf{x} = (I_H, I_V)^\top$ and separating new infections from all other transitions,
\begin{equation*}
\mathcal{F} = \begin{pmatrix} b_H \beta_{VH} \dfrac{I_V}{N_V}\, S_H \\[10pt] b_V \beta_{HV} \dfrac{I_H}{N_H}\, S_V \end{pmatrix}, \qquad \mathcal{V} = \begin{pmatrix} \gamma I_H \\[6pt] \mu I_V \end{pmatrix}.
\end{equation*}
The Jacobians evaluated at the DFE are
{\setlength{\arraycolsep}{1pt}
\begin{equation*}
F = \begin{pmatrix} 0 & \displaystyle b_H \beta_{VH} \frac{N_H}{N_V} \\[10pt] \displaystyle b_V \beta_{HV} \frac{N_V}{N_H} & 0 \end{pmatrix}, ~~ V = \begin{pmatrix} \gamma & 0 \\[4pt] 0 & \mu \end{pmatrix}, ~~ FV^{-1} = \begin{pmatrix} 0 & \displaystyle\frac{b_H \beta_{VH}}{\mu}\frac{N_H}{N_V} \\[12pt] \displaystyle\frac{b_V \beta_{HV}}{\gamma}\frac{N_V}{N_H} & 0 \end{pmatrix}.
\end{equation*}}
\paragraph{Basic reproduction number} Since $FV^{-1}$ is $2\times 2$ block off-diagonal, its eigenvalues are $\pm\sqrt{a_{12}a_{21}}$, and the population-ratio factors $N_H/N_V$ and $N_V/N_H$ cancel exactly in the product, giving
\begin{equation}\label{eq:R0_vb}
\mathcal{R}_0 = \rho(FV^{-1}) = \sqrt{\frac{b_V \beta_{VH}}{\mu} \cdot \frac{b_H \beta_{HV}}{\gamma}} = \sqrt{\mathcal{R}_{VH} \cdot \mathcal{R}_{HV}},
\end{equation}
where the \emph{one-way reproduction numbers} are
\begin{equation*}\label{eq:one_way_vb}
\mathcal{R}_{VH} = \frac{b_V \beta_{VH}}{\mu}, \qquad \mathcal{R}_{HV} = \frac{b_H \beta_{HV}}{\gamma}.
\end{equation*}
Here $\mathcal{R}_{VH}$ counts the expected new human infections generated by one infectious vector: it delivers $b_V$ bites per unit time, each transmitting to a susceptible human with probability $\beta_{VH}$, over a vector lifespan of $1/\mu$. Symmetrically, $\mathcal{R}_{HV}$ counts the expected new vector infections generated by one infectious human: it receives $b_H$ bites per unit time, each acquiring infection with probability $\beta_{HV}$, over a human infectious period of $1/\gamma$.

\paragraph{$\mathcal{R}_0$ structure and disease transmission mechanism} The geometric mean form of $\mathcal{R}_0$ is a direct consequence of the bipartite transmission structure: a single transmission cycle requires both a vector-to-host step and a host-to-vector step; and a completed cycle produces $\mathcal{R}_{VH}\cdot\mathcal{R}_{HV}$ secondary cases. Thus, the averaged per-step (generation) secondary cases, $\mathcal{R}_0$, is obtained by $\mathcal{R}_0^2 = \mathcal{R}_{VH}\cdot\mathcal{R}_{HV}$, which leads to the geometric mean in \cref{eq:R0_vb}.

\subsubsection{Heterosexual Sexually-transmitted Disease}
\label{sec:std}
Following Hethcote and Yorke \cite{HethcoteYorke1984}, let $S_F$, $I_F$ and $S_M$, $I_M$ denote the numbers of susceptible and infected females and males, with fixed total sizes $N_F$ and $N_M$. Each sex is characterized by an encounter rate ($a_F$, $a_M$: average daily contacts per individual with the opposite sex), a per-encounter transmission probability ($\beta_{MF}$, $\beta_{FM}$), and a mean infectious duration ($d_F$, $d_M$). The forces of infection are
\begin{equation}\label{eq:foi_std}
\lambda_F = a_F \beta_{MF} \frac{I_M}{N_M}, \qquad \lambda_M = a_M \beta_{FM} \frac{I_F}{N_F},
\end{equation}
which are structurally identical to~\cref{eq:foi_vb} with females as hosts and males as vectors. The SIS dynamics are
\begin{align*}
\frac{dS_F}{dt} &= -\lambda_F S_F + \frac{I_F}{d_F}, \qquad \frac{dI_F}{dt} = \lambda_F S_F - \frac{I_F}{d_F}, \\
\frac{dS_M}{dt} &= -\lambda_M S_M + \frac{I_M}{d_M}, \qquad \frac{dI_M}{dt} = \lambda_M S_M - \frac{I_M}{d_M}.
\end{align*}
As noted explicitly by Hethcote and Yorke \cite{HethcoteYorke1984}, the STD model above is \emph{``formally the same as a host-vector model.''}

Applying the NGM method (the intermediate steps omitted given the structural correspondence with the vector-borne model) gives
\begin{equation*}
\mathcal{R}_0 = \sqrt{a_M \beta_{MF} d_M \cdot a_F \beta_{FM}d_F} = \sqrt{\mathcal{R}_{MF}\cdot\mathcal{R}_{FM}},
\end{equation*}
where $\mathcal{R}_{MF} = a_M \beta_{MF} d_M$ and $\mathcal{R}_{FM} = a_F \beta_{FM} d_F$ are the sex-specific one-way reproduction numbers: $\mathcal{R}_{MF}$ counts the expected new female infections generated by one infectious male over his infectious period ($M \to F$), and $\mathcal{R}_{FM}$ the reverse ($F \to M$). The reappearance of the geometric mean confirms that this form is a signature of bipartite topology rather than a feature specific to vector-borne disease.

\section{Vertical Transmission}\label{sec:vertical}
In vertical transmission, the ``infected'' element passes from parent to offspring through reproduction \cite{busenberg1993vertically}. $\mathcal{R}_0$ here measures the invasion fitness of the infected type relative to the resident, that is, a fitness ratio rather than a transmission count, while retaining $\mathcal{R}_0 = 1$ as the invasion threshold from a small introduction of infection.

\subsection{\W Population Replacement: $\mathcal{R}_0$ as Population Reproduction Ratio}\label{sec:wolbachia}
\textit{Wolbachia pipientis} is an intracellular endosymbiont carried naturally by a wide range of arthropod species and has been demonstrated to suppress the vectorial capacity of \textit{Aedes aegypti} for dengue, Zika, Chikungunya, and several other arboviruses \cite{walker2011wmel,hoffmann2011successful}. Unlike the horizontally transmitted pathogens considered in \cref{sec:horizontal}, \W spreads primarily through vertical (maternal) transmission: infected females pass the bacterium to a fraction $v_w$ of their offspring. A second mechanism, cytoplasmic incompatibility (CI), renders matings between uninfected females and \Wns-infected males embryonically lethal, conferring a reproductive advantage on infected females. Because \W infection also imposes fitness costs via reduced fecundity ($\phi_w \leq \phi_u$), elevated adult mortality ($\mu_{fw} \geq \mu_{fu}$), and elevated egg mortality ($\mu_{ew} \geq \mu_{eu}$), a small introduced cohort of \Wns-infected mosquitoes typically cannot invade, and successful population replacement requires an initial release exceeding a critical threshold \cite{qu2018modeling}. The central modeling objective is therefore to characterize this threshold in terms of the per-generation fitness of the infected and uninfected cohorts.

\begin{figure}[ht!]
\centering
\includegraphics[width=0.65\linewidth]{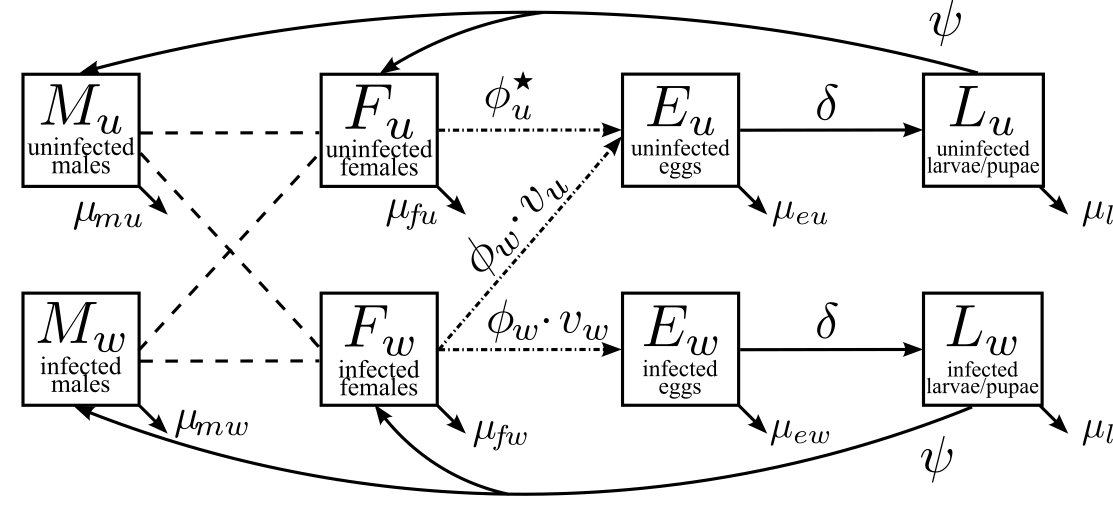}
\caption{Transmission diagram for the \W population replacement model. Solid arrows denote demographic transitions through the egg ($E$), larval ($L$), female ($F$), and male ($M$) life stages for uninfected (subscript $u$) and \Wns-infected (subscript $w$) cohorts. The dashed arrow denotes vertical transmission from infected females to offspring.}
\label{fig:Wolbachia}
\end{figure}
We consider a multistage ODE model from \cite{florez2023wolbachia} that tracks the mosquito population across life stages and both infection statuses (\cref{fig:Wolbachia}): uninfected and \Wns-infected eggs ($E_u$, $E_w$), larvae ($L_u$, $L_w$), adult females ($F_u$, $F_w$), and adult males ($M_u$, $M_w$). Eggs hatch to larvae at per-capita rate $\delta$ and suffer mortality at per-capita rate $\mu_{ej}$ ($j \in \{u,w\}$); larvae mature to adults at per-capita rate $\psi$ with density-independent mortality rate $\mu_l$. The larval stage further suffers density-dependent mortality through competition for shared resources, modeled via a logistic growth term with carrying capacity $K_l$; maturation produces adult females with probability $b_f$ and adult males with probability $b_m = 1 - b_f$. Uninfected females lay eggs at a per-capita rate $\phi_u$, but CI renders matings with \Wns-infected males sterile, so only those females pairing with an uninfected male, with probability $M_u/(M_u + M_w)$ under random mating, produce viable eggs. Infected females are unaffected by CI, lay eggs at per-capita rate $\phi_w$, and transmit \W maternally to a fraction $v_w$ of their offspring; a fraction $v_u= 1 - v_w$ of offspring from infected females are uninfected, reflecting potential imperfect maternal transmission. The governing system is
\begin{align*}
\frac{dE_u}{dt} &= \phi_u \frac{M_u}{M_u+M_w} F_u + v_u\,\phi_w F_w - (\delta+\mu_{eu})\,E_u, & \frac{dE_w}{dt} &= v_w\,\phi_w F_w - (\delta+\mu_{ew})\,E_w, \notag \\
\frac{dL_u}{dt} &= \delta\!\left(1-\frac{L_u+L_w}{K_l}\right)\!E_u - (\psi+\mu_l)\,L_u, & \frac{dL_w}{dt} &= \delta\!\left(\!1-\frac{L_u+L_w}{K_l}\!\right)\!E_w \!- \!(\psi+\mu_l)\,L_w, \label{eq:wolbachia_odes} \\
\frac{dF_u}{dt} &= b_f\psi\,L_u - \mu_{fu}\,F_u, & \frac{dF_w}{dt} &= b_f\psi\,L_w - \mu_{fw}\,F_w, \notag \\
\frac{dM_u}{dt} &= b_m\psi\,L_u - \mu_{mu}\,M_u, & \frac{dM_w}{dt} &= b_m\psi\,L_w - \mu_{mw}\,M_w. \notag
\end{align*}

\paragraph{$\mathcal{F}$--$\mathcal{V}$ decomposition} 
We treat \Wns-infected individuals as the ``infected'' population. The DFE consists of a fully uninfected population, $E_w = L_w = F_w = M_w = 0$, with
\begin{equation*}
F_u^* = b_f\frac{\psi}{\mu_{fu}}\,L_u^*, \quad M_u^* = b_m\frac{\psi}{\mu_{mu}}\,L_u^*, \quad E_u^* = b_f \frac{\psi}{\mu_{fu}} \frac{\phi_u}{\delta + \mu_{eu}} L_u^*, \quad L_u^* = K_l\!\left(1 - \frac{1}{\mathcal{G}_{0u}}\right),
\end{equation*}
where $\mathcal{G}_{0u}$ is the uninfected population reproduction number defined in \cref{eq:G0_wolbachia}; the DFE is biologically feasible if and only if $\mathcal{G}_{0u} > 1$. Taking infected state vector $\mathbf{x} = (E_w,\, L_w,\, F_w,\, M_w)^\top$ and separating new-infection terms from transitions, we have 
\begin{equation*}
\mathcal{F} = \begin{pmatrix} v_w\phi_w F_w \\ 0 \\ 0 \\ 0 \end{pmatrix}, \qquad 
\mathcal{V} = 
\begin{pmatrix} (\delta+\mu_{ew})\,E_w \\ -\delta\!\left(1-\frac{L_u+L_w}{K_l}\right)\!E_w + (\psi+\mu_l)\,L_w \\ -b_f\psi\,L_w + \mu_{fw}\,F_w \\ -b_m\psi\,L_w + \mu_{mw}\,M_w \end{pmatrix}.
\end{equation*}
Linearizing at the DFE, the Jacobian matrices are
\begin{equation*}
F = \begin{pmatrix} 0 & 0 & v_w\phi_w & 0 \\ 0 & 0 & 0 & 0 \\ 0 & 0 & 0 & 0 \\ 0 & 0 & 0 & 0 \end{pmatrix}, \qquad V = \begin{pmatrix} \delta+\mu_{ew} & 0 & 0 & 0 \\ -\delta/\mathcal{G}_{0u} & \psi+\mu_l & 0 & 0 \\ 0 & -b_f\psi & \mu_{fw} & 0 \\ 0 & -b_m\psi & 0 & \mu_{mw} \end{pmatrix}.
\end{equation*}

\paragraph{Basic reproduction number}
This gives the basic reproduction number
\begin{equation}\label{eq:R0_wolbachia}
\mathcal{R}_0 = \rho(FV^{-1}) = v_w \cdot \frac{\mu_{fu}\,\phi_w\,(\delta+\mu_{eu})}{\mu_{fw}\,\phi_u\,(\delta+\mu_{ew})} =\frac{\mathcal{G}_{0w}}{\mathcal{G}_{0u}},
\end{equation}
where the \emph{population reproduction numbers} for the \Wns-infected and uninfected cohorts are
\begin{equation}\label{eq:G0_wolbachia}
\mathcal{G}_{0w} = v_w\,b_f \cdot \frac{\delta}{\delta+\mu_{ew}} \cdot \frac{\psi}{\psi+\mu_l} \cdot \frac{\phi_w}{\mu_{fw}}, \qquad \mathcal{G}_{0u} = b_f \cdot \frac{\delta}{\delta+\mu_{eu}} \cdot \frac{\psi}{\psi+\mu_l} \cdot \frac{\phi_u}{\mu_{fu}}.
\end{equation}
Each quantity $\mathcal{G}_{0j}$ ($j \in \{u,w\}$) is a product of stage-specific survival and fecundity factors: $\delta/(\delta+\mu_{ej})$ is the probability that an egg survives to hatch; $\psi/(\psi+\mu_l)$ is the probability that a larva survives to adulthood; and $b_f\phi_j/\mu_{fj}$ is the expected number of female offspring produced per adult female over her lifetime. The additional factor $v_w$ in $\mathcal{G}_{0w}$ is the maternal transmission rate, the probability that an offspring of an infected female inherits \Wns. Notably, the common larval survival factor $\psi/(\psi+\mu_l)$ and the sex ratio $b_f$ cancel in the ratio $\mathcal{G}_{0w}/\mathcal{G}_{0u}$, so $\mathcal{R}_0$ depends only on the infection-induced differentials in egg mortality ($\mu_{ej}$), fecundity ($\phi_j$), and adult female longevity ($1/\mu_{fj}$), together with $v_w$.

\paragraph{$\mathcal{R}_0$ structure and disease transmission mechanism}
The ratio form $\mathcal{R}_0 = \mathcal{G}_{0w}/\mathcal{G}_{0u}$ is the defining feature of the vertical-transmission setting and is fundamentally different from horizontal transmission. In horizontal transmission, the susceptible host population is slowly varying (assuming a large population size) over the infection timescale near the DFE, so counting the absolute number of secondary infections produced by a single infectious individual is a meaningful threshold to determine if the pathogen can spread.

In vertical transmission, by contrast, both the uninfected and infected cohorts are simultaneously reproducing populations. An absolute reproduction count for the infected cohort alone does not determine invasion; it must instead be compared against the uninfected cohort's reproduction: invasion occurs when the fraction of infection in the population increases. The ratio form $\mathcal{G}_{0w}/\mathcal{G}_{0u}$ serves this purpose by normalizing the infected cohort's raw reproductive output by the resident's own reproductive output. The ratio structure directly reflects the competitive relationship between the two cohorts inherent to vertical transmission.

\subsection{\textit{Medea} Gene Drive: $\mathcal{R}_0$ as a Genotypic Fitness Ratio} \label{sec:medea}
\textit{Medea} (Maternal-Effect Dominant Embryonic Arrest) is a threshold-dependent gene drive acting through a maternal-effect toxin--antidote mechanism: a female carrying the drive deposits a toxin during oogenesis, and her offspring die unless they inherit the drive-linked antidote. Coupled with a disease-refractory trait, it offers an alternative population-replacement strategy to the \W strategy of \cref{sec:wolbachia}, and serves as a prototype for the broader family of toxin-antidote gene drives that followed. In \cite{qu2026modeling}, an ODE compartmental model is proposed to track mosquito abundance by genotypes: wild-type ($++$), heterozygous ($M+$), and \textit{Medea}-homozygous ($MM$), indexed by $i=0,1,2$. In contrast to the \W model, which describes a competition between two (infected and uninfected) cohorts, the \textit{Medea} model leads to a competition among three genotypes. 

\begin{figure}[htbp]
\centering
\includegraphics[width=\linewidth]{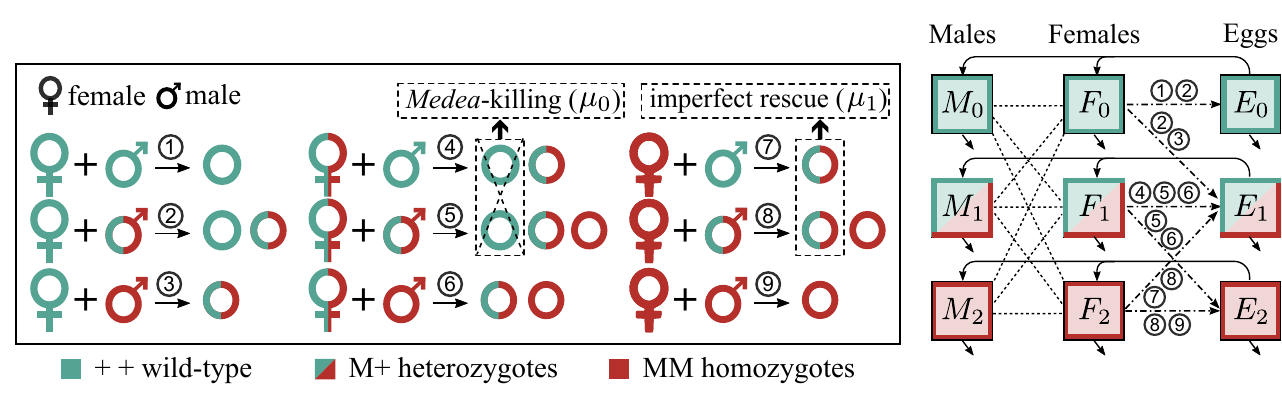}
\caption{Left: Transmission of \textit{Medea} gene drive depends on parental genotypes. Nine possible crosses between female and male genotypes, numbered \circled{1}--\circled{9}, following Mendelian inheritance but modified for the \textit{Medea}-killing (toxin) and imperfect rescue. More details in the text. Right: Schematic representation for complex transmission routes of the \textit{Medea} gene drive across life stages and genotypes. Subscript 0 = wild-type (++); 1 = heterozygous (M+); 2 = \textit{Medea} homozygotes (MM).}
\label{fig:medea}
\end{figure}

The state variables are the juvenile ($E_i$, combining eggs, larvae, and pupae), adult female ($F_i$), and adult male ($M_i$) populations of each genotype. Under homogeneous mixing and equal mating competence across male genotypes, the cross $F_i \times M_j$ contributes juveniles at rate $F_i(M_j/N_M)\phi_{ij}$ with $N_M = M_0+M_1+M_2$; juveniles develop at per-capita rate $\psi$ and mature in a $1{:}1$ sex ratio; the mortality rates $\mu_{ei},\mu_{fi},\mu_{mi}$ and the fecundities $\phi_{ij}$ are genotype-specific and carry the fitness cost of the drive; and density dependence enters through a logistic factor $(1-N_E/K)$ on juvenile recruitment, with $N_E = E_0+E_1+E_2$. Two \textit{Medea}-specific parameters account for the toxin-antidote mechanism: the killing leakage $\mu_0$ is the fraction of wild-type offspring of $M+$ mothers that survive the maternal toxin (crosses $\circled{4}$, $\circled{5}$), and the rescue efficiency $\mu_1$ is the fraction of heterozygous offspring of $MM$ mothers that survive its double dose (crosses $\circled{7}$, $\circled{8}$), with $\mu_0 = 0$, $\mu_1 = 1$ the perfect \textit{Medea}-killing and perfect rescue, respectively. The governing system is
\begin{align*}
\frac{dE_i}{dt} &= \Lambda_i\!\left(1-\frac{N_E}{K}\right) - (\psi+\mu_{ei})\,E_i, \qquad i = 0,1,2,  \\
\frac{dF_i}{dt} &= \frac{1}{2}\psi E_i - \mu_{fi}F_i, \qquad i = 0,1,2,  \\
\frac{dM_i}{dt} &= \frac{1}{2}\psi E_i - \mu_{mi}M_i, \qquad i = 0,1,2, 
\end{align*}
where the recruitment terms, $\Lambda_i (i = 0,1,2)$, assembled from the nine crosses of \cref{fig:medea} under Mendelian segregation modified by the drive, are
\begin{align*}
\Lambda_0 &= \frac{1}{N_M}\Bigl[F_0M_0\phi_{00} + \frac12 F_0M_1\phi_{01} + \mu_0\bigl(\frac12 F_1M_0\phi_{10} + \frac14 F_1M_1\phi_{11}\bigr)\Bigr], \\
\Lambda_1 &= \frac{1}{N_M}\Bigl[\frac12 F_0M_1\phi_{01} + F_0M_2\phi_{02} + \frac12 F_1M_0\phi_{10} + \frac12 F_1M_1\phi_{11} + \frac12 F_1M_2\phi_{12} \\
&\qquad + \mu_1\bigl(F_2M_0\phi_{20} + \frac12 F_2M_1\phi_{21}\bigr)\Bigr],\\
\Lambda_2 &= \frac{1}{N_M}\Bigl[\frac14 F_1M_1\phi_{11} + \frac12 F_1M_2\phi_{12} + \frac12 F_2M_1\phi_{21} + F_2M_2\phi_{22}\Bigr].
\end{align*}
\paragraph{$\mathcal{F}$--$\mathcal{V}$ decomposition} We treat the \textit{Medea}-carrying stages as the ``infected'' population. The DFE consists entirely of wild-type mosquitoes,
\begin{equation}\label{eq:medea_mfe}
E_0^\ast = \left(1-\frac{1}{\mathcal{G}_{00}}\right)K, \quad F_0^\ast = \frac{\psi}{2\mu_{f0}}E_0^\ast, \quad M_0^\ast = \frac{\psi}{2\mu_{m0}}E_0^\ast, 
\end{equation}
with $E_1^\ast = E_2^\ast = F_1^\ast = F_2^\ast = M_1^\ast = M_2^\ast = 0$.
The $\mathcal{G}_{00}$ is the wild-type population reproduction number, given in \cref{eq:G_medea}. Taking the infected state vector $\mathbf{x} = (E_1,F_1,M_1,E_2,F_2,M_2)^\top$, and separating the new infection terms from transitions,
\begin{equation*}
\mathcal{F} = \begin{pmatrix} \Lambda_1\left(1-\frac{N_E}{K}\right) \\ 0 \\ 0 \\ \Lambda_2\left(1-\frac{N_E}{K}\right) \\0 \\ 0 \end{pmatrix}, \qquad \mathcal{V} = \begin{pmatrix} (\psi+\mu_{e1})\,E_1 \\\mu_{f1}F_1 - \frac{1}{2}\psi E_1\\ \mu_{m1}M_1 - \frac{1}{2}\psi E_1 \\(\psi+\mu_{e2})\,E_2 \\ \mu_{f2}F_2 - \frac{1}{2}\psi E_2 \\\mu_{m2}M_2 - \frac{1}{2}\psi E_2 \end{pmatrix}.
\end{equation*}
Linearizing at the DFE, the Jacobian matrices are
\begin{equation*}
F = \begin{pmatrix} 0 & \vartheta\,\mu_{f0}\,\phi_{10} & \vartheta\,\mu_{m0}\,\phi_{01} & 0 & 2\vartheta\,\mu_1\,\mu_{f0}\,\phi_{20} & 2\vartheta\,\mu_{m0}\,\phi_{02} \\ 0 & \cdots &\cdots & \cdots& \cdots& 0 \\ \vdots & & & & & \vdots \\ 0 & \cdots & \cdots & \cdots &\cdots & 0 \end{pmatrix}, \quad \vartheta = \dfrac{\mu_{e0}+\psi}{\phi_{00}\,\psi},
\end{equation*}
\begin{equation*}
V = \begin{pmatrix} V_1 & O_{3\times3} \\ O_{3\times3} & V_2 \end{pmatrix}, \quad V_i = \begin{pmatrix} \psi+\mu_{ei} & 0 & 0 \\ -\frac12\psi & \mu_{fi} & 0 \\ -\frac12\psi & 0 & \mu_{mi} \end{pmatrix}.
\end{equation*}
\paragraph{Basic reproduction number} This gives the basic reproduction number
\begin{equation}\label{eq:R0_medea}
\mathcal{R}_0 \!=\! \rho(FV^{-1}) \!= \!\frac{\bigl(\mu_{f1}\mu_{m0}\phi_{01} + \mu_{f0}\mu_{m1}\phi_{10}\bigr)(\mu_{e0}+\psi)}{2\,\mu_{f1}\mu_{m1}\,\phi_{00}\,(\mu_{e1}+\psi)} = \frac{\mathcal{G}_{011}\,\frac{1}{\mu_{m1}} + \mathcal{G}_{101}\,\frac{1}{\mu_{m0}}}{\mathcal{G}_{00}\,\frac{1}{\mu_{m0}}},
\end{equation}
where the \emph{population reproduction numbers} are
\begin{equation}\label{eq:G_medea}
\mathcal{G}_{00} = \frac{1}{2}\cdot\frac{\psi}{\psi+\mu_{e0}}\cdot\frac{\phi_{00}}{\mu_{f0}}, \quad \mathcal{G}_{011} = \frac{1}{4}\cdot\frac{\psi}{\psi+\mu_{e1}}\cdot\frac{\phi_{01}}{\mu_{f0}}, \quad \mathcal{G}_{101} = \frac{1}{4}\cdot\frac{\psi}{\psi+\mu_{e1}}\cdot\frac{\phi_{10}}{\mu_{f1}}.
\end{equation}
Each $\mathcal{G}_{ijk}$ counts the expected number of genotype-$k$ female offspring produced over the lifetime of a genotype-$i$ female mated to a genotype-$j$ male, and they share the same stage-structure as the \W population reproduction numbers in \cref{eq:G0_wolbachia}; $\mathcal{G}_{00}$ describes the resident cross $F_0\times M_0$ (1/2 for half of the offspring is female), while $\mathcal{G}_{011}$ and $\mathcal{G}_{101}$ describe the reciprocal heterozygous crosses $F_0\times M_1$ and $F_1\times M_0$, each carrying a coefficient $1/4$ because Mendelian segregation assigns the $M+$ genotype to only half of the offspring, a half of which is female. In the absent any fitness cost ($\mu_{e0}=\mu_{e1}$, $\mu_{f0}=\mu_{f1}$, $\mu_{m0}=\mu_{m1}$, $\phi_{00}=\phi_{01}=\phi_{10}$), we have $\mathcal{G}_{011} = \mathcal{G}_{101} = \frac12\mathcal{G}_{00}$, thus, $\mathcal{R}_0 = 1$ exactly.

\paragraph{$\mathcal{R}_0$ structure and disease transmission mechanism} As in the \W model, $\mathcal{R}_0$ takes a ratio form, reflecting the vertical transmission mechanism and the competition in per-generation reproduction between the introduced type and the resident type. However, the \textit{Medea} ratio \cref{eq:R0_medea} differs from the \W ratio \cref{eq:R0_wolbachia} in three aspects. 
First, only two of the three genotypes appear in the \textit{Medea} $\mathcal{R}_0$: M+ in the numerator and ++ in the denominator, and the third $MM$ genotype is absent. This is because producing an MM offspring requires a \textit{Medea} allele from both parents, making it a second-order process that is invisible to the linearization at the DFE. Second, the reproduction of the M+ cohorts (the numerator) sums over the two reciprocal heterozygous crosses rather than a single term, because the \textit{Medea} allele can be passed through either parental line ($F_0\times M_1$ or $F_1\times M_0$), unlike the strictly maternal transmission of \Wns. 

Lastly, each reproduction number is weighted by a mean male lifespan factor, $1/\mu_{m1}$ for $\mathcal{G}_{011}$ ($F_0\times M_1$), and $1/\mu_{m0}$ for $\mathcal{G}_{101}$ ($F_1\times M_0$) and $\mathcal{G}_{00}$ ($F_0\times M_0$), capturing the impact of fitness cost in reduced male survival. However, no such factor appears in the \W ratio \cref{eq:R0_wolbachia}. This reflects a difference in the role males play in the two transmission mechanisms. In the \W case, males play only an auxiliary role in reproduction, and they do not pass \W infection to their offspring, directly contributing to the spread of infection. In the \textit{Medea} case, by contrast, both parents can transmit it to their offspring, so both male and female lifespans enter the population reproduction numbers.

% -----------------------------------------------------------------------
\section{Discussion and Conclusion}
\label{sec:conclusion}
This review surveys a range of compartmental models, spanning distinct transmission features, each of which admits a characteristic algebraic structure for the basic reproduction number $\mathcal{R}_0$. We summarize these structural forms, together with their biological mechanisms, in \cref{tab:taxonomy}. We demonstrate that the functional form of $\mathcal{R}_0$ varies systematically with the underlying transmission mechanism. 

This structural diversity is biologically meaningful: although $\mathcal{R}_0$ is most commonly used simply as a threshold governing disease outbreak or invasion, its algebraic structure provides insights into disease transmission and control that the numerical value would not convey on its own:
%Row-by-row commentary
\begin{itemize}
\item The weighted-sum structure, $\mathcal{R}_0 = \sum_i p_i\,\mathcal{R}_0^{(i)}$, arises from the SP and DI models. It gives an explicit breakdown of contributions from each infectious stage, which depends on both the stage-specific reproduction number and the corresponding entry probability. A highly infectious minority stage (large stage $\mathcal{R}_0^{(i)}$ but small entry probability $p_i$) does not necessarily contribute more infections than a moderately infectious majority stage (moderate stage $\mathcal{R}_0^{(i)}$ but large entry probability $p_i$). This breakdown, therefore, identifies the stage that drives transmission and the most efficient target for intervention.

\item The maximum structure, $\mathcal{R}_0 = \max_i\{\mathcal{R}_0^{(i)}\}$, arises from multi-strain or multi-patch models. The global $\mathcal{R}_0$ is dominated by the single largest strain or patch-level $\mathcal{R}_0^{(i)}$. Targeting that dominant strain or patch first would be more efficient than evenly distributing, though achieving global elimination requires every strain or patch to be suppressed.

\item The geometric-series (cyclic) structure, $ \mathcal{R}_0 = \mathcal{R}_0^{\mathrm{cycle}}/(1-P_\gamma) = \mathcal{R}_0^{\mathrm{cycle}} \sum_{i=0}^\infty {(P_\gamma)^i}$, arises from models with disease relapses. Recognizing this decomposition separates the per-cycle reproduction number, $\mathcal{R}_0^{\mathrm{cycle}}$, from the relapse probability, $P_\gamma$, two quantities that would otherwise remain entangled within the raw algebraic expression for $\mathcal{R}_0$. This separation reveals that a relapse probability close to unity can substantially amplify global transmission, so that the global $\mathcal{R}_0$ may exceed one even when the per-cycle reproduction number itself remains below one.

\item The geometric mean structure, $\mathcal{R}_0 =\sqrt{\mathcal{R}_{HV}\,\mathcal{R}_{VH}}$, arises from bipartite transmission, such as vector-borne diseases or heterosexual STDs. A complete transmission cycle involves a two-step process, and separating the one-way reproduction numbers, $\mathcal{R}_{HV}$ and $\mathcal{R}_{VH}$, for each direction helps to measure the contribution from each side. Due to the multiplicative form, control efforts can therefore be directed at whichever host class is more logistically or economically tractable to target.

\item Fitness-ratio structure, $\mathcal{R}_0 = \mathcal{G}_{0w}/\mathcal{G}_{0u}$, arises from models with vertical transmission, reflecting the competition in per-generation reproduction between the introduced cohort, $\mathcal{G}_{0w}$, and the resident cohort, $\mathcal{G}_{0u}$. Such a fitness-ratio structure motivates constructing the population reproduction numbers for the competing cohorts near DFE, which allows explicit observation of how the demographic differences between them, such as infection-induced fitness costs, enter the ratio and impact the invasion dynamics near the DFE.
\end{itemize}

The taxonomy in \cref{tab:taxonomy} is not an exhaustive list of the forms that $\mathcal{R}_0$ can take across compartmental models; however, we expect that some of the underlying principles discussed here can be transferred to settings not covered in this review. For example, models with environmental or indirect transmission, such as cholera \cite{tien2010multiplea,mukandavire2011estimating}, typically yield an $\mathcal{R}_0$ that consists of direct and environmental components, $\mathcal{R}_0 = \mathcal{R}_0^{\mathrm{direct}} + \mathcal{R}_0^{\mathrm{env}}$. This generalizes the weighted-sum structure of \cref{sec:DISP} by summing over distinct horizontal transmission routes. 
In age-structured PDE models \cite{qu2023modeling,inaba1990threshold}, the discrete weighted sum is generalized to a continuous integral over the age dimension, $\mathcal{R}_0 = \int_a \mathcal{R}_{0,a} P(a) da$, where the age-specific reproduction number $\mathcal{R}_{0,a}$ is weighted by the age distribution $P(a)$. We also do not treat more complex model classes, such as multiscale or spatially structured models.

The numerical value of $\mathcal{R}_0$ is only half the story; its algebraic structure is the other half. Interpreting its structure alongside the value itself is essential to fully understand the mechanisms of disease transmission and identify effective points of control. This review is intended as a research synthesis, consolidating existing $\mathcal{R}_0$ results under a common structural taxonomy, and a pedagogical reference for those new to the field. It could also serve as a practical starting point for researchers encountering an unfamiliar transmission mechanism.

\section*{Acknowledgments}
Z.Q. was supported by the National Science Foundation under award DMS-2316242. A.A. conducted this research as a participant in the Research Science Scholars Program at Saint Mary's Hall, and thanks the program for its support. 

\bibliographystyle{plain}
\bibliography{references}

\appendix

\section{Derivation step for iNTS Model}\label{sec:app_nts}
The Jacobian matrices for the infected subsystem are given by 
\begin{equation*}
F = \frac{\partial \mathcal{F}}{\partial \mathbf{x}}=
\begin{pmatrix}
D_\mathcal{F}^1 & D_\mathcal{F}^1 & D_\mathcal{F}^1 & D_\mathcal{F}^1 \\[0.2em]
D_\mathcal{F}^2 & D_\mathcal{F}^2 & D_\mathcal{F}^2 & D_\mathcal{F}^2 \\[0.2em]
D_\mathcal{F}^3 & D_\mathcal{F}^3 & D_\mathcal{F}^3 & D_\mathcal{F}^3 \\[0.2em]
D_\mathcal{F}^4 & D_\mathcal{F}^4 & D_\mathcal{F}^4 & D_\mathcal{F}^4 
\end{pmatrix}
\quad \text{and} \quad
V=\frac{\partial \mathcal{V}}{\partial \mathbf{x}}=
\begin{pmatrix}
D_\mathcal{V}^1 & & & \\
 & D_\mathcal{V}^2 & & \\
 & & D_\mathcal{V}^3 & \\
 & & & D_\mathcal{V}^4 
\end{pmatrix}~,
\end{equation*}
where 
\begin{equation*}
D_\mathcal{F}^\kappa =
\begin{pmatrix}
\varphi^\kappa & \varphi^\kappa & \varphi^\kappa & 0 & \varphi^\kappa \\
0&0&0&0&0 \\
0&0&0&0&0 \\
0&0&0&0&0 \\
0&0&0&0&0
\end{pmatrix},
\qquad
\varphi^\kappa = c\beta^\kappa \frac{S_0^\kappa}{N_0}, \qquad N_0 = \sum_{\kappa=1}^4 S_0^\kappa,
\end{equation*}
{\setlength{\arraycolsep}{-3pt}
\begin{equation*}
D_\mathcal{V}^\kappa =
\begin{pmatrix}
\gamma_{12}^\kappa+\gamma_{\scriptscriptstyle 1H}^\kappa+\mu^\kappa & 0 & 0 & 0 & -\gamma_{\scriptscriptstyle A1}^\kappa \\
-\gamma_{\scriptscriptstyle 1H}^\kappa & \gamma_{\scriptscriptstyle H2}^\kappa+\gamma_{\scriptscriptstyle HA}^\kappa+\gamma_{\scriptscriptstyle HR}^\kappa+\mu^\kappa & 0 & 0 & 0 \\
-\gamma_{12}^\kappa & -\gamma_{\scriptscriptstyle H2}^\kappa & \gamma_{\scriptscriptstyle 2D}^\kappa+\gamma_{\scriptscriptstyle 2M}^\kappa+\mu^\kappa & 0 & 0 \\
0 & 0 & -\gamma_{\scriptscriptstyle 2M}^\kappa & \gamma_{\scriptscriptstyle MA}^\kappa+\gamma_{\scriptscriptstyle MD}^\kappa+\gamma_{\scriptscriptstyle MR}^\kappa+\mu^\kappa & 0 \\
0 & -\gamma_{\scriptscriptstyle HA}^\kappa & 0 & -\gamma_{\scriptscriptstyle MA}^\kappa & \gamma_{\scriptscriptstyle A1}^\kappa+\mu^\kappa
\end{pmatrix}.
\end{equation*}}

\end{document}